\documentstyle[epsfig]{elsart}
\def\Journal#1#2#3#4{{#1} {#2} (#4) #3.}
\def\NIMR{Nucl. Instr. and Meth. in Phys. Res. A}
\def\NPA{Nucl. Phys. A}
\def\PLB{Phys. Lett.  B}
\def\PRL{Phys. Rev. Lett.}
\def\PRC{Phys. Rev. C}
\def\PR{Phys. Rep.} 
\def\ZPA{Z. Phys. A}
\def\MPL{Mod. Phys. Lett. A}
\def\EPJ{Eur. Phys. J. A}
\def\be{\begin{equation}}
\def\ee{\end{equation}}

\begin{document}
\begin{frontmatter}

\title{Transition from in-plane to out-of-plane azimuthal enhancement 
in Au+Au collisions}

\author[buc,dar]{A.~Andronic\thanksref{infos}}, 
\author[buc]{G.~Stoicea},
\author[buc]{M.~Petrovici}, 
\author[buc]{V.~Simion}, 
\author[cle]{P.~Crochet}, 
\author[cle]{J.~P.~Alard}, 
\author[dar]{R.~Averbeck},
\author[cle]{V.~Barret},
\author[zag]{Z.~Basrak}, 
\author[cle]{N.~Bastid}, 
\author[cle]{A.~Bendarag},
\author[bud]{G.~Berek},
\author[zag]{R.~\v{C}aplar}, 
\author[dar]{A.~Devismes}, 
\author[cle]{P.~Dupieux}, 
\author[zag]{M.~D\v{z}elalija},
\author[hdu]{M.~Eskef},
\author[dar]{Ch.~Finck},
\author[bud]{Z.~Fodor},
\author[dar]{A.~Gobbi},
\author[mos]{Y.~Grishkin},
\author[dar]{O.~N.~Hartmann},
\author[hdu]{N.~Herrmann}, 
\author[dar]{K.~D.~Hildenbrand},
\author[kor]{B.~Hong},
\author[bud]{J.~Kecskemeti}, 
\author[kor]{Y.~J.~Kim}, 
\author[war]{M.~Kirejczyk}, 
\author[zag]{M.~Korolija},
\author[dre]{R.~Kotte}, 
\author[dar]{T.~Kress}, 
\author[dar]{R.~Kutsche},
\author[mos]{A.~Lebedev}, 
\author[kor]{K.~S.~Lee},
\author[hdu]{Y.~Leifels}, 
\author[kur]{V.~Manko}, 
\author[hdu]{H.~Merlitz},
\author[dre]{W.~Neubert},
\author[hdu]{D.~Pelte}, 
\author[dre]{C.~Plettner},
\author[str]{F.~Rami}, 
\author[dar]{W.~Reisdorf},
\author[str]{B.~de~Schauenburg},
\author[dar]{D.~Sch\" ull}, 
\author[bud]{Z.~Seres}, 
\author[war]{B.~Sikora}, 
\author[kor]{K.~S.~Sim}, 
\author[war]{K.~Siwek-Wilczy\'nska}, 
\author[mos]{V.~Smolyankin}
\author[hdu]{M.~R.~Stockmeier}, 
\author[kur]{M.~Vasiliev},
\author[str]{P.~Wagner}, 
\author[war]{K.~Wi\'sniewski}, 
\author[dre]{D.~Wohlfarth},
\author[kur]{I.~Yushmanov},
\author[mos]{A.~Zhilin}

\collab{FOPI Collaboration}

\address[buc]{National Institute for Physics and Nuclear Engineering, 
              Bucharest, Romania}
\address[bud]{KFKI Research Institute for Particle and Nuclear Physics, Budapest, Hungary}
\address[cle]{Laboratoire de Physique Corpusculaire, IN2P3/CNRS, 
              and Universit\'{e} Blaise Pascal, Clermont-Ferrand, France}
\address[dar]{Gesellschaft f\"ur Schwerionenforschung, Darmstadt, Germany}
\address[dre]{Forschungszentrum Rossendorf, Dresden, Germany}
\address[hdu]{Physikalisches Institut der Universit\"at Heidelberg, 
              Heidelberg, Germany}
\address[mos]{Institute for Theoretical and Experimental Physics, 
              Moscow, Russia}
\address[kur]{Kurchatov Institute, Moscow, Russia}
\address[kor]{Korea University, Seoul, Korea}
\address[str]{Institut de Recherches Subatomiques,IN2P3-CNRS and 
              Universit\'{e} Louis Pasteur, Strasbourg, France}
\address[war]{Institute of Experimental Physics, Warsaw University, Poland}
\address[zag]{Rudjer Boskovic Institute, Zagreb, Croatia}

\thanks[infos]{Corresponding author: GSI, Planckstr. 1, D-64291 Darmstadt,
Germany; Email:~A.Andronic@gsi.de; Phone: +49 615971 2769; 
Fax: +49 615971 2989}

\begin{abstract}

The incident energy at which the azimuthal distributions in semi-central heavy
ion collisions change from in-plane to out-of-plane enhancement -- 
$E_{tran}$ is studied as a function of mass of emitted particles, their 
transverse momentum and centrality for Au+Au collisions.
The analysis is performed in a reference frame rotated with the sidewards 
flow angle ($\Theta_{flow}$) relative to the beam axis.

A systematic decrease of $E_{tran}$ as function of 
mass of the reaction products, their transverse momentum
and collision centrality is evidenced. 

The predictions of a microscopic transport model (IQMD) are
compared with the experimental results.
\end{abstract}


\end{frontmatter}

\begin{flushleft}\it{PACS: 25.75.Ld;25.70.Pq}\end{flushleft}

\section{Introduction}

A decade ago, two types of azimuthal anisotropies have been evidenced 
in heavy ion collisions \cite{wil1,gut1,dem,gut2}. 
Following their shape relative to the
reaction plane and the main mechanism behind them they have been baptized as 
in-plane and out-of-plane enhancement or rotational-like \cite{wil1} and
squeeze-out \cite{gut1,dem,gut2} phenomena, respectively.
How the transition from one to the other scenario takes place as a function 
of incident energy, for different collision geometries and types of the 
reaction products was a question addressed few years later for a light 
\cite{pop} and heavy system \cite{but1}.
Soon, it was realized that a detailed study of such a transition from in-plane
to out-of-plane emission as a function of incident energy could give more
insight into the relative contribution of the attractive and repulsive forces,
the lifetime of the emitting source, its rotational energy and expansion 
dynamics. Therefore many studies concentrated on this subject 
\cite{wil2,tsa1,aa1,but2,bas,cro1,she,zhe}.

Below this transition energy, $E_{tran}$, the experimental data show an 
in-plane enhancement of the azimuthal distribution.
Theoretical calculations established that
a rotating compound system created by the mean field, dissipating angular 
momenta and excitation energy via particle emission, stays at the origin of 
this observation \cite{wil2,gar}.

Increasing the beam energy the mean field which contributes to the formation of
a rotating compound system becomes less important and processes based on
nucleon-nucleon interaction start to be predominant. The hydrodynamic
approaches used to predict different phenomena which take place
in the relativistic regime of heavy ion collisions have shown that 
a preferential emission of the ``squeezed'' participant zone in the 
free phase space, not hindered by the spectator nuclear matter, takes place.
The out-of-plane emission of the nuclear matter has been called squeeze-out
\cite{sto} and has been confirmed experimentally by Plastic Ball 
\cite{gut1,gut2} and Diogene \cite{dem} Collaborations few years later. 
Afterwards, the generality of this phenomenon was evidenced experimentally.
Thus, pions \cite{bri1,ven}, kaons \cite{sen}, neutrons \cite{lei,lam}, 
light particles \cite{bri2,wan}, proton-like particles \cite{tsa1} 
and intermediate mass fragments \cite{bas,cro1} show a similar squeeze-out 
pattern in the azimuthal distributions.
Detailed theoretical investigations of this phenomenon were also undertaken,
using microscopic transport models \cite{wel,aic,har1,har2,kis,bas1,bas2}.

The existence of an azimuthally symmetric flow evidenced in central
heavy ion collisions \cite{jeo,hsi,pet1,rei} can be regarded as the 
extreme case of the out-of-plane flow. 
In other words, the squeeze-out is the result of the expansion of the hot and 
compressed participant zone in the presence of the projectile and target 
spectators.
Although it is obvious that the cleanest signal on the collective 
expansion of hot and compressed baryonic matter can be obtained from central 
collisions, one has to accept that this is an extreme case and the study of a 
rotating hot and compressed object remains an appealing subject.
If the angular momenta and the shadowing realized by the spectator matter which
come into the game bring more complications at first glance, they can be used 
as internal clocks to gain more information on the expansion dynamics.

The incident energy where the effects of these three competing processes, 
expansion, rotation and shadowing, compensate each other could be looked as a 
benchmark observable for deeper understanding of the expansion dynamics.

In this paper we present the results of a multidimensional analysis of the 
transition energy as a function of the mass of emitted particles, their 
transverse momentum and for different collision centralities.
Early FOPI studies \cite{but1,bas,cro1}, done using the experimental data from 
Phase I of the detector, have been continued with the data obtained 
using the full phase space coverage of Phase II of this device 
\cite{aa1,aa2,pet2}.
Relative to the Phase I FOPI studies, the present data make possible the 
azimuthal distributions studies in a reference frame with the $z$ axis along 
the sidewards flow direction, with no upper limitation on transverse momentum 
and as function of the mass of light charged particles. 

Chapter 2 describes in more details the experiment and the experimental 
data analysis. 
The experimental results are presented in chapter 3. Chapter 4 is dedicated to 
the comparison of the results with the predictions of a microscopic transport 
model. Conclusions are presented in chapter 5.

\section{Experimental Details}

\subsection{Setup} 

The present experimental configuration \cite{rit,pel} was used to study 
Au+Au, Xe+CsI and Ni+Ni collisions at 90, 120, 150, 250 and 400$\cdot$A MeV,
aiming to continue our studies started with Phase I of the FOPI \cite{gob}
detector with much better phase space coverage and as a function of the 
baryonic number involved in the process. The beam energies are the 
mid-target energies, taking into account the upstream energy loss of 
the beam. The results presented in this paper refer to the Au+Au system.

Since details of the first phase can be found in our earlier publications,
especially in ref. \cite{gob}, we describe here the main features of the 
Central Drift Chamber (CDC), the main extension of Phase I of the FOPI facility
\cite{rit,pel} used in this experiment. The CDC is a drift chamber that 
performs the tracking of the path followed by all charged reaction products 
emitted in the polar angular range 33$^\circ <\theta_{lab}<$150$^\circ$. 
The path being curved in the magnetic field of a superconducting solenoid, 
one can use the transverse momentum and specific energy loss for mass 
identification of the reaction products. Full azimuthal coverage is realized 
by this tracking device. The detector is subdivided in 16 sectors, each 
containing 60 sense wires and 60 potential wires, all aligned parallel to the 
beam axis. The sense wires are resistive and readout on both ends, which gives
the possibility to reconstruct the position along the wire of the hit
via charge division. The reconstruction of the initial position of the ionized
cell along the ionizing particle path is obtained making use of the electrons
drift velocity of about 43.7 $\mu$m/ns and of the Lorentz angle
$\alpha_{L}$=32$^\circ$. As the direction of 
the drift electrons cannot be measured, each hit has a symmetric partner 
relative to the sense wires plane (mirror hit). They combine and form mirror 
tracks. In order to have a criterion to distinguish between real and mirror 
tracks, the sense wires planes are tilted 8$^\circ$ relative to the radial 
direction starting from the beam axis. Within this geometry the mirror tracks 
do not originate anymore from the target. Furthermore, the sense wires within 
a plane are alternatively staggered by $\pm$100~$\mu$m, which could be also 
used to reject the mirrored tracks. The individual signal wires of 50~$\mu$m 
thick NiCr-based alloy  with a resistance of 500~$\Omega$/m vary between 86 
and 190~cm. The mixture used for the chamber consists of 88\% Ar, 10\% 
isobuthane and 2\% CH$_4$ at a slight overpressure.

The CDC was operated in a ``split-mode'' in order to increase its dynamical 
range. While the most outer 30 potential wires were operated at a nominal 
voltage of -1.55 kV, optimum for a good resolution for energetic light 
particles (Z=1,2), the most inner 30 potential wires had a lower voltage of 
-1.1 kV, such that the amplitude of the signals corresponding to highly 
ionizing fragments (Z=3-6) to be within the dynamical range of our sampling 
convertors, Flash ADCs (8 bit, non-linear).
The relative momentum resolution $\sigma_{p_{t}}/p_{t}$ varies from 4\% 
for $p_{t}<$~0.5~GeV/c to about 12\% for $p_{t}$=2~GeV/c.
A helium bag placed between the target and the forward subdetector of Phase I
\cite{gob}, following 
the conical shape of the CDC forward endcap, was used in order to decrease the 
energy thresholds for the intermediate mass fragments flying in the 
forward direction at $\Theta_{lab}<$30$^\circ$. 
It is worth to mention here that within the phase space covered by the
Phase I subdetector the reaction products are identified by their charge using 
specific energy loss and time-of-flight information, while the fragments 
detected by the CDC are identified by their mass, as the information on the 
specific energy loss and magnetic rigidity was available.

\begin{figure}[hbt]
\centering\mbox{\epsfig{file=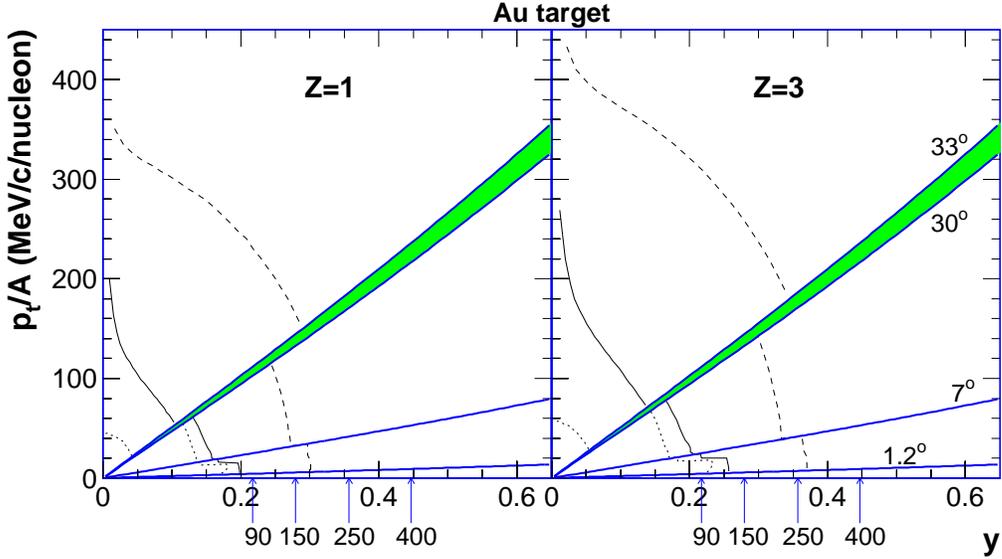, width=0.95\textwidth}}
\caption{FOPI Phase II acceptance as a function of the transverse momentum per 
nucleon ($p_t/A$) and laboratory reference frame rapidity for Z=1 and Z=3 
reaction products. The c.m. rapidities corresponding to different incident 
energies are marked by arrows.}
\label{fig-1}
\end{figure}

The phase space coverage of the whole device used in the present experiment,
in a transverse momentum - $p_t$ versus rapidity - $y$ representation, 
can be followed in Fig.~\ref{fig-1}.
The  borders between different subdetectors at 1.2$^\circ$, 7$^\circ$, 
30$^\circ$ and 33$^\circ$ are represented by thick lines. 
The shadowed area between 30$^\circ$ and 33$^\circ$ corresponds to the 
region between the Plastic Wall (PW) and CDC which is not covered at all.
The low thresholds (LT) are represented by thin continuous lines. They 
correspond to the measured values (dotted lines) after correction for the
energy loss in the media before the active detectors.
This was done assuming that the interaction has taken place in the middle 
of the target.
For polar angles smaller than 30$^\circ$ the low threshold corresponds
to the momenta for which the particles reach the PW in order to deliver 
a time signal while the specific energy loss is obtained from a layer of 
ionization chambers or thin plastic scintillators in front of the 
PW \cite{gob}.
For polar angles larger than 30$^\circ$ LT was considered to correspond 
to the momenta for which the ionizing particles could reach the outer 
radius of the CDC.

The dashed lines correspond, for the forward detector, to the momenta at 
which the plastic scintilators are just penetrated by the corresponding
particles. In this case the energy loss correction is negligible.
For the CDC the dashed lines correspond to the momenta at which the 
energy loss corrections are below 1\%. 

\subsection{Analysis techniques} 
The combination of the very high hit densities into tracks associated to the
true particles is a serious challenge of a tracking algorithm 
\cite{pel,herr}.
The data analyzed in the present paper have been obtained using for the CDC a 
local tracking method - ``track-following'' - in a version developed within 
the FOPI collaboration \cite{herr}.

\begin{figure}[htb] 
\hspace{-0.mm}\mbox{\epsfig{file=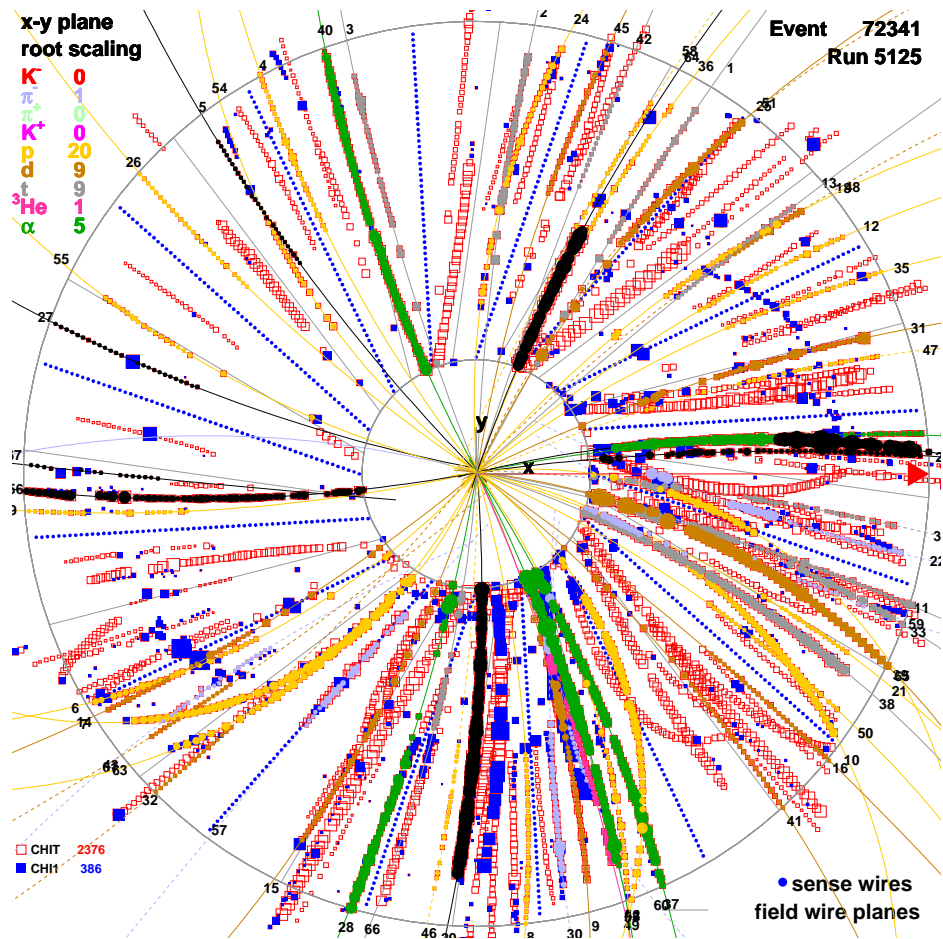,width=0.65\textwidth}}
\caption{CDC $xy$ plane cross section for a Au+Au, E=250$\cdot$A MeV central 
event.} \label{fig-2}
\end{figure}

Fig.~\ref{fig-2} shows a representation where the performance 
of the algorithm in track resolution can be followed. It corresponds to a 
projection in a plane perpendicular to the beam axis of the trajectories 
associated to the charged particles seen by the CDC in a typical event from 
central Au+Au collisions at 250$\cdot$ AMeV.
The straight lines of dots are representing the anode (sense) wires, while 
the rest of symbols are the registered hits. Grey symbols are belonging to 
reconstructed tracks and the open squares are their mirrored counterparts.
The dark full squares are the leftovers remained unassigned to any track.
 
\begin{figure}[hbt]
\begin{tabular}{lr} \begin{minipage}[t]{0.48\textwidth}
\centering\mbox{\epsfig{file=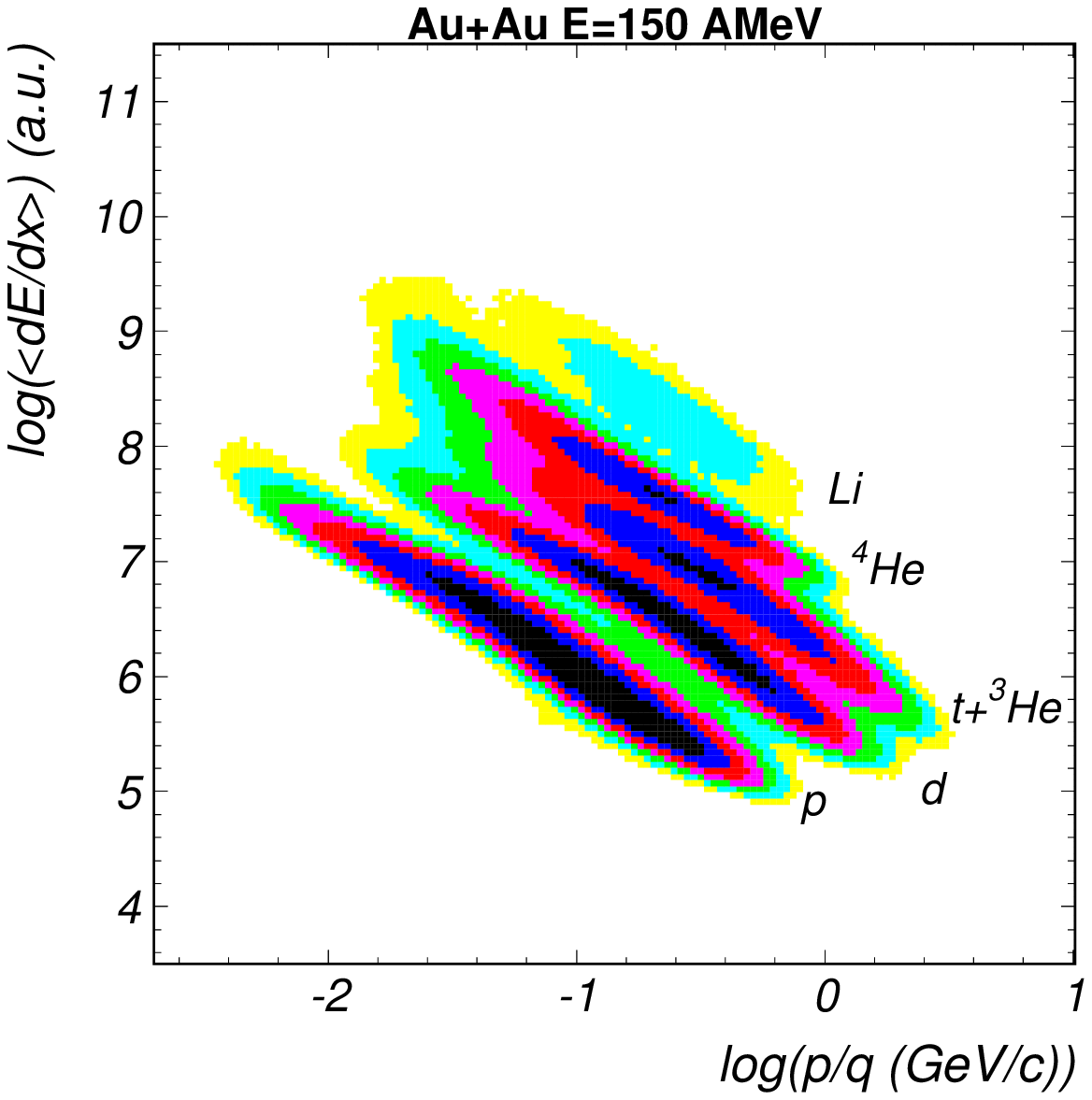,width=0.96\textwidth}}
\caption
{The mass identification performance of CDC. The scale on $z$ is logarithmic.} 
\label{fig-3} \end{minipage} &
\begin{minipage}[t]{0.48\textwidth} 
\centering\mbox{\epsfig{file=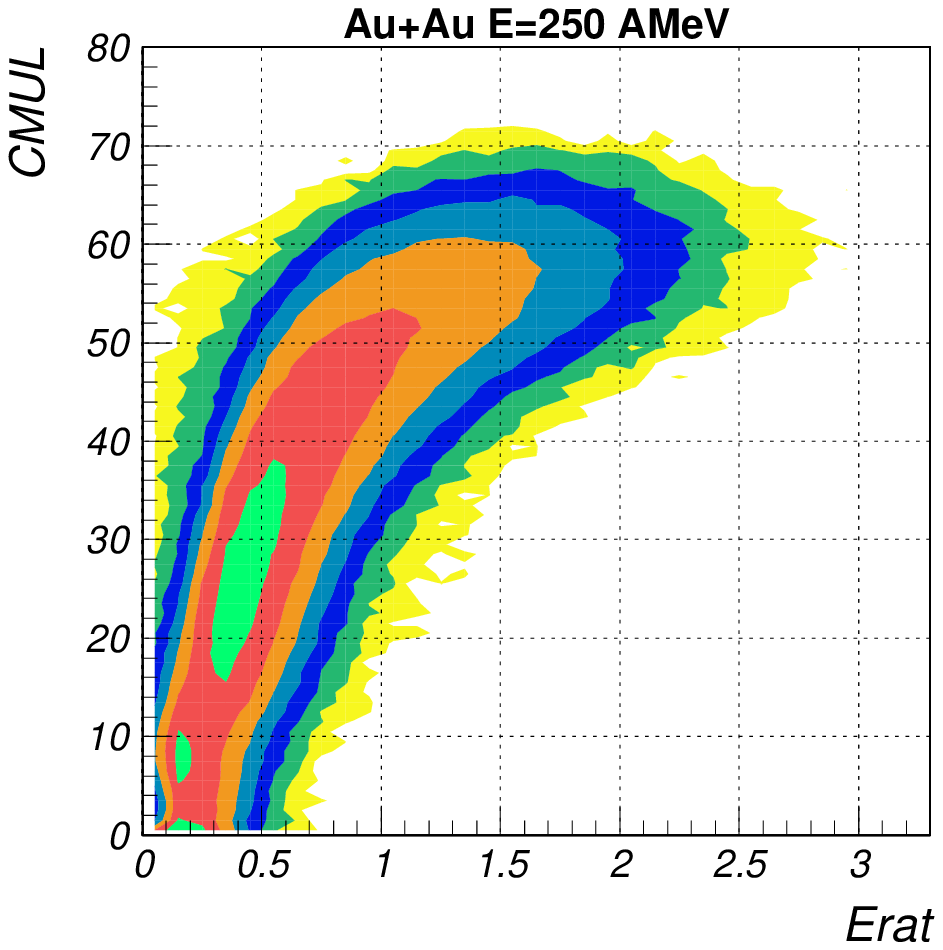, width=0.97\textwidth}}
\caption{CMUL-$Erat$ correlation for Au+Au, E=250$\cdot$AMeV.} \label{fig-4}
\end{minipage} \end{tabular} 
\end{figure}

The mass identification performance of CDC using specific energy loss 
($dE/dx$) versus transverse momentum divided by charge ($p_t/q$) of the 
detected particles can be followed in Fig.~\ref{fig-3}.
A cross-check of the present calibration of CDC \cite{aa3} was done using 
comparison with our earlier data measured with Si-CsI telescopes \cite{pog}. 
Good agreement between the two sets of data, in spectral shape and 
yields, was obtained.
 However, using only the CDC information, the A=3 branch contains a mixed 
contribution from $^3$He and tritium (t) fragments (as can be seen in Fig.3).
Consequently, the momentum associated to $^{3}$He fragments will be 
underestimated by a factor of two if $q$=1, corresponding to hydrogen isotopes,
is considered. Based on the phase space distribution of $^{3}$He and t 
obtained in one of our previous experiments \cite{pog}, the contribution
of $^{3}$He to A=3 yield for a given $p_{t}^{(0)}$ range (using $q$=1) is 
smaller that 15\%.
As it can be seen later, this amount of contamination does not influence 
different systematics of A=3 fragments. They nicely fall between those 
corresponding to A=2 and A=4 fragments.   

Most of the events were registered under ``medium bias'' (MB) trigger, 
corresponding to different values for the outer Plastic Wall multiplicity 
as a function of energy and mass of the colliding systems, such to reduce the 
background contribution written on tape, estimated from target free runs 
to be below 5\%.

Fig.~\ref{fig-4} shows the multiplicity distribution of the reaction products 
detected and identified by CDC, CMUL, as function of 
$E{rat}=\sum_{i}E_{\perp,i}/\sum_{i}E_{\parallel,i}$ (the sums run over
all products detected in an event) for the energy of 250$\cdot$A MeV 
under hardware ``medium bias'' trigger. As expected, besides a clear 
correlation between these two observables, producing the ridge of the 
displayed distribution, strong fluctuations are visible. The trends observed 
in this representation are in good qualitative agreement with the microscopic 
transport model prediction, particularly the IQMD version \cite{har3}, 
extensively used for the present analysis. 
At large impact parameters (low CMUL and $E{rat}$ values) the particle 
multiplicity has a higher selectivity of the collision geometry, while 
towards higher centrality (larger CMUL and $E{rat}$ values) the
$Erat$ does better. Following similar recipe as the one used in Phase I for
PMUL \cite{her2}, a CMUL selection was devised in order to select the collision
geometry. 

\begin{figure}[hbt]
\centering\mbox{\epsfig{file=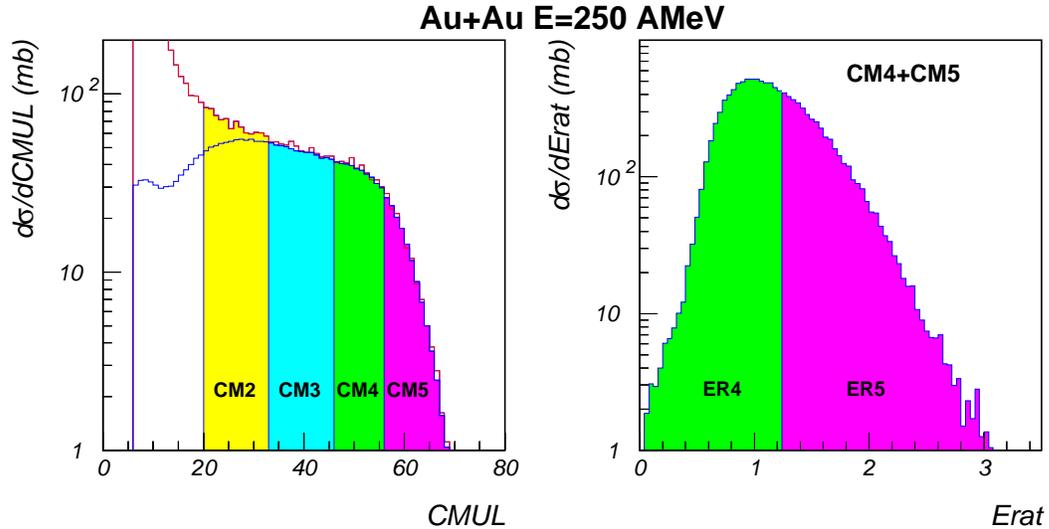, width=0.99\textwidth}}
\caption{CMUL and $Erat$ distributions for 250$\cdot$A MeV.
Different centrality bins used in the present analysis can be followed.} 
\label{fig-5}
\end{figure}

In Fig.~\ref{fig-5} we show the CMUL distribution (both for minimum and medium
bias triggers) and the $Erat$ distribution for CM4+CM5 bins. 
The CM2 and CM3 regions have been used in order to select impact parameters in
the range of 6--8 fm and 4--6 fm, respectively and ER4 and ER5 for 2--4 fm and 
0--2 fm, respectively. 
The impact parameter values have been obtained  using a geometrical 
``sharp cut-off'' approximation for the reaction cross section. 

\begin{figure}[hbt]
\centering\mbox{\epsfig{file=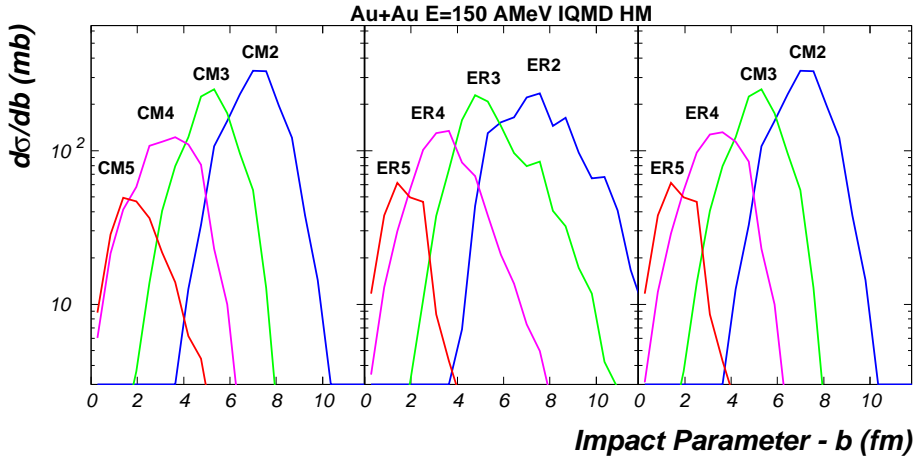, width=.98\textwidth}}
\caption{The impact parameter distributions extracted from the IQMD (HM) model
at the energy of 150$\cdot$A MeV for the experimental selection criteria using
CMUL (CM) or $Erat$ (ER) observables.} \label{fig-21}
\end{figure}

To check our centrality selection method we used events generated with the 
IQMD model (see Section \ref{s:model}), for which the impact parameter is 
known. Fig.~\ref{fig-21} shows the distribution of the impact parameters
for different centralities selected in the way 
outlined above, for IQMD HM, Au+Au at 150$\cdot$A MeV. 
As one could see, the argument for our centrality selection 
based on the CMUL-$Erat$ correlation is confirmed by the model predictions.

\subsection{Reaction plane and flow angle determination} 
Besides the collision geometry, the reaction plane and the sidewards flow 
angle determination are crucial for the quality of the information 
extracted from azimuthal distributions. 
The transverse momentum analysis method has been used for the reaction plane 
reconstruction \cite{dan}. In order to avoid the autocorrelations, the 
particle of interest was not included in the reaction plane determination. 
In such a situation the momentum conservation is violated and a recoil 
correction can be done. One has to mention that such a correction brings back 
the autocorrelations and, taking into account that we will concentrate on 
heavy systems, characterized by high particle multiplicity, where the recoil 
correction is negligible, we rather preferred not to use it.

\begin{figure}[hbt]                  
\centering\mbox{\epsfig{file=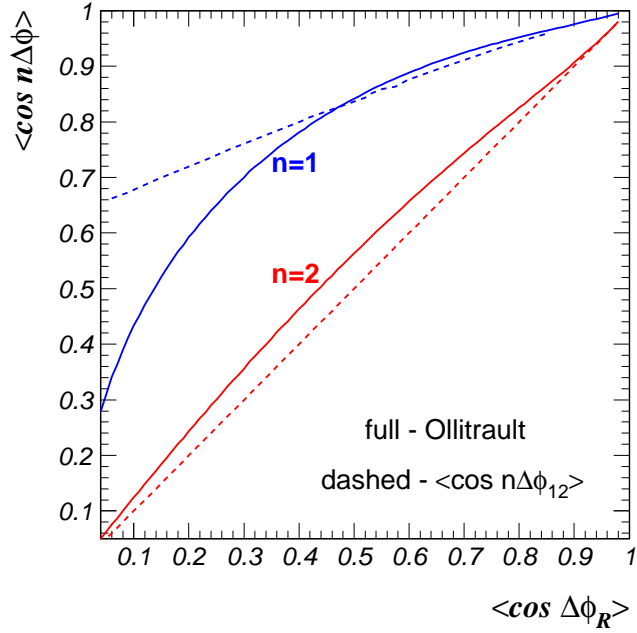, width=0.6\textwidth}}
\caption{The correction factors for $a_{1}$ and $a_{2}$ (see Eq. \ref{eq-2} 
in the text) due to the reaction plane resolution.} \label{fig-6}
\end{figure}

The correction for the reaction plane fluctuations due to the finite 
multiplicity and detector biases has been estimated using the method proposed
by Ollitrault \cite{oli}.
This is based as the previous one \cite{dan} on a random subdivision of each 
event in two and calculating the difference between the azimuth angles of the 
reaction planes extracted from these subevents: 
$\Delta\phi_R = \Phi_1 - \Phi_2$.
The dependence of this correction on the cosine of $\Delta\phi_R$ is 
represented in Fig.~\ref{fig-6}. 
In the same figure is represented the correction using 
$\langle\cos n\Delta\Phi_{12}\rangle$, where $\Delta\Phi_{12}$=$\Delta\phi_R$/2
is a measure of the reaction plane resolution \cite{dan}.
Table~\ref{tab-1} contains the
$\langle\cos\Delta\phi_R\rangle$ values for the energies and centralities
studied in the present paper.

\begin{table}[hbt]
\begin{center}
\caption{Experimental values of $\langle\cos\Delta\phi_R\rangle$ for the 
measured energies and centrality bins.} \label{tab-1}
\begin{tabular}{cccccc} \hline
Beam  Energy &90 & 120 & 150 & 250 & 400 \\ \hline
CM2&~~0.151 &~~0.390 & ~~0.454 & ~~0.756 & ~~0.820 \\ \hline
CM3&~~0.252 &~~0.520 & ~~0.644 & ~~0.827 & ~~0.864 \\ \hline
ER4&~~0.283 &~~0.527 & ~~0.679 & ~~0.797 & ~~0.817 \\ \hline
ER5&~~0.164 &~~0.326 & ~~0.471 & ~~0.563 & ~~0.600 \\ \hline
\end{tabular}
\end{center}
\end{table}

\begin{figure}[h]
\centering\mbox{\epsfig{file=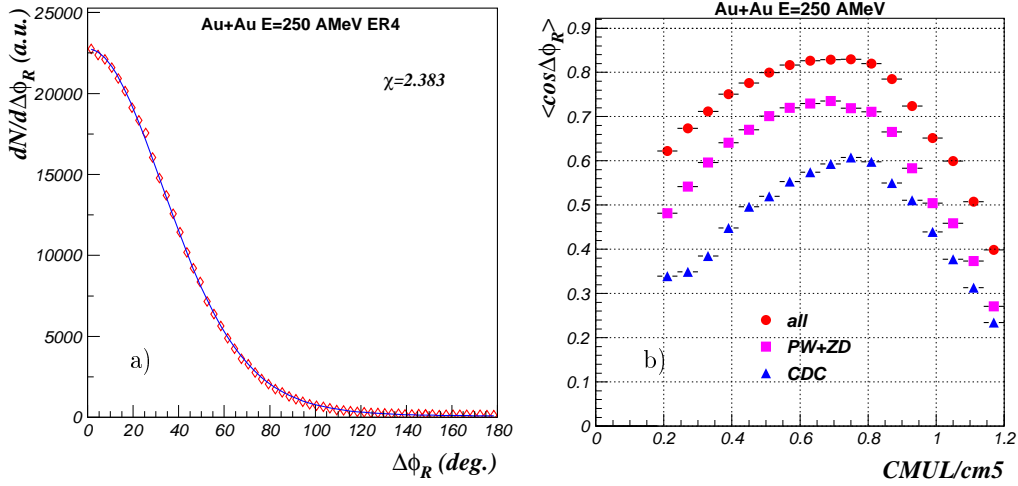, width=1.02\textwidth}}
\caption{ a) Reaction plane resolution for E=250$\cdot$A MeV, ER4; b)
$\langle\cos\Delta\phi_R\rangle$ dependence on the scaled multiplicity 
($cm5$ is the lower limit for the CM5 bin; $cm5$=56 for the present example).}
\label{fig-7}
\end{figure}

An example of $\Delta\phi_R$
distribution for ER4 centrality is presented in Fig.~\ref{fig-7}a for
E=250$\cdot$A MeV. Such experimental representations 
are used for determining the dimensionless parameter $\chi$ 
using the following expression (Eq.~12 in ref.~\cite{oli}):

\begin{equation}
\frac{dN}{d\Delta\phi_R}=\frac{{e}^{-\chi^2/2}}{2}
\left\{\frac{2}{\pi}(1+\chi^2/2)+z[I_0(z)+\mathbf{L}_0(z)]+\chi^2/2[I_1(z)+
\mathbf{L}_1(z)]\right\}
\end{equation}
Where z=$\chi^2_{I}$cos$\Delta\phi_R$ and $\mathbf{L}_0$, $\mathbf{L}_1$
are the modified Struve functions.
The line in Fig.~\ref{fig-7}a is the result of a fit using the above equation.
The correction factors are then given by Eq. 8 in ref. \cite{oli}:
\begin{equation}
\langle\cos n\Delta\phi\rangle = \frac{\sqrt{\pi}}{2}\chi
{e}^{-\chi^2/2}\left[I_{\frac{n-1}{2}}\left(\frac{\chi^2}{2}\right)+
I_{\frac{n+1}{2}}\left(\frac{\chi^2}{2}\right)\right]
\end{equation}
where $I_{k}$ is the modified Bessel function of order $k$.

A representation of the $\langle\cos\Delta\phi_R\rangle$ as a 
function of scaled CMUL for the energy of 250$\cdot$A MeV is presented 
in Fig.~\ref{fig-7}b.
Table~\ref{tab-1} shows clearly the influence of the incident energy
and of the impact parameter  on the precision of the 
reaction plane determination. For a given incident energy the 
precision reaches a maximum for an intermediate value of the impact parameter 
(multiplicity) as it can be seen also in Fig.~\ref{fig-7}b. This figure 
shows also the behaviour of $\langle\cos\Delta\phi_R\rangle$ as a function of 
CMUL using the  information delivered by the full device or only by some of 
its subdetectors. In the subsequent analysis, the full detector has been used
for the reaction plane reconstruction.

Already in the first paper in which the experimental confirmation of the 
out-of-plane enhancement was reported \cite{gut1} it was stressed the 
importance of performing the azimuthal distribution analysis in a reference
system which has the polar axis along the sidewards flow direction 
\cite{gut2}. There are many recipes to determine the flow angle. 
The methods based on the sphericity analysis \cite{gyu} or on fitting the 
momentum distribution of the particles detected in the participant region by a 
three dimensional anisotropic Gaussian distribution \cite{gos} may be affected
by the contributions coming from the spectator components.
Information on flow angle can be also obtained using sidewards flow analysis
\cite{ram}.
In the present analysis we use a different procedure for theta flow 
determination \cite{wan,dan2,aa1}.
The flow angle $\Theta_{flow}$ is defined as being the angle with which all 
the events in a given centrality class have been rotated in order to maximize 
the squeeze-out signal or by carefully analyzing the $p_x$ - y distributions
in the region of $E_{tran}$ value. 

\begin{figure}[htb]
\centering\mbox{\epsfig{file=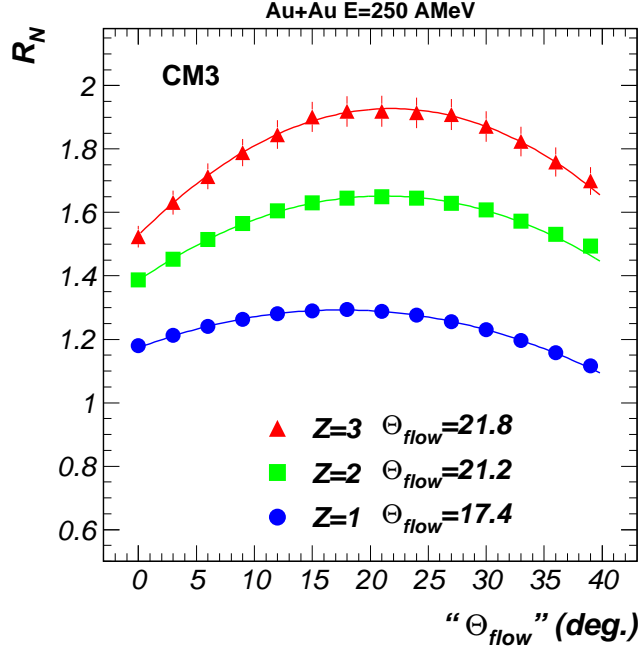, width=0.6\textwidth}}
\caption{Squeeze-out ratio $R_N$ in a reference system rotated in the 
reaction plane by ``$\Theta_{flow}$'' angle relative to the beam axis, 
integrated over all momenta.}
\label{fig-8}
\end{figure}

\begin{figure}[htb]
\centering\mbox{\epsfig{file=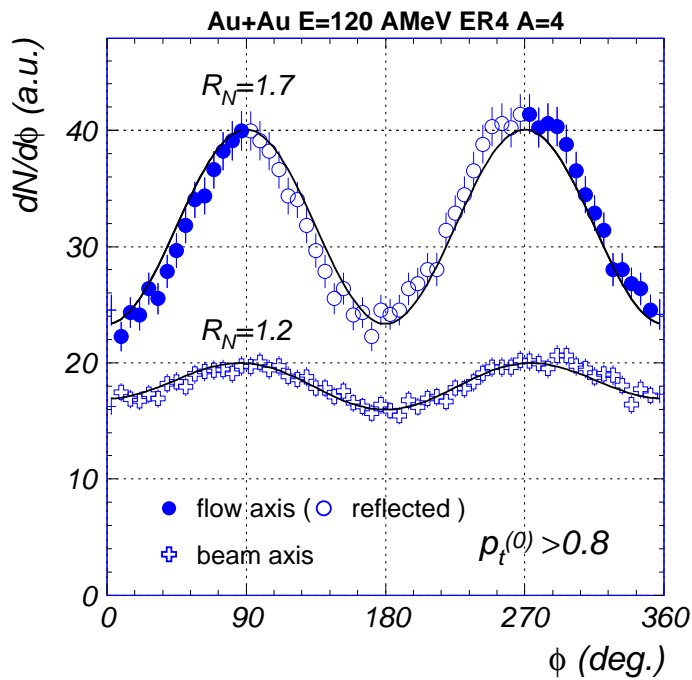, width=0.65\textwidth}}
\caption{Azimuthal distributions for A=4 particles in the beam and in the 
rotated reference frame, 80$^\circ\leq\Theta_{cm}\leq$100$^\circ$.} 
\label{fig-11}
\end{figure}

Fig.~\ref{fig-8} shows the dependence of $R_N^{exp}$: 
\begin{equation}
R_N^{exp}=\frac{\frac{dN}{d\phi}(\phi=90^o)+\frac{dN}{d\phi}(\phi=270^o)}
{\frac{dN}{d\phi}(\phi=0^o)+\frac{dN}{d\phi}(\phi=180^o)}
\end{equation}
estimated in a coordinate system which has two axes ($x-z$) in the reaction 
plane and the polar axis $z$ at an angle ``$\Theta_{flow}$'' relative to the 
beam axis, as function of the rotation angle ``$\Theta_{flow}$'' for 
250$\cdot$A MeV incident energy and CM3 centrality bin. 
As one could see, the maximum value of $R_N$ and the width of its distribution
as a function of rotation angle ``$\Theta_{flow}$'' shows a 
dependence on the mass of the analyzed particle. 
The mass dependence of the distribution width is a direct consequence 
of the sidewards flow profile.
Part of this coming from thermal fluctuations, it is natural that 
with increasing the mass of the analyzed particle the distribution becomes 
narrower \cite{ram}.
If $\Theta_{flow}$ extracted from a standard sidewards flow analysis 
is affected by the flow profile \cite{ram}, the mean value of $R_{N}$ 
distribution as a function of ``$\Theta_{flow}$'' does not depend on it.
Nevertheless, the data still show a mass dependence of $\Theta_{flow}$.
It could be attributed to the dynamics of the collision. 
The light particles quite probable are originating from the initial 
stage of the collision, when the sidewards flow is not yet developed and 
transparency effects are more important, leading to lower values of 
the sidewards flow angle.
On the contrary, heavier fragments could originate 
from later stages of the collision, when the sidewards flow has developed.
Part of this pattern could be also due to preequilibrium and sequential 
emission processes. Already for Z=2 and Z=3 fragments the difference is 
negligible. Consequently, the $\Theta_{flow}$ values used in the present 
analysis correspond to these reaction products.
In order to see the difference between the azimuthal distributions in the 
rotated reference system along the sidewards flow direction and in the 
non-rotated one, along the beam axis, we represent in Fig.~\ref{fig-11}
the corresponding two azimuthal distributions for the energy of
120$\cdot$A MeV.
We have selected here particles of mass A=4, the centrality bin ER4 and 
normalized transverse momentum per nucleon to the projectile momentum per 
nucleon in the c.m. system $p_t^{(0)}=(p_t/A)/(p_P^{cm}/A_P)>$0.8. 

The effect of the rotated reference frame on the $R_N$ values 
makes unnecessary any other argument for the need of carrying out these types 
of studies in this reference system.

\subsection{The influence of the detector acceptance} 
\begin{figure}[htb]
\centering\mbox{\epsfig{file=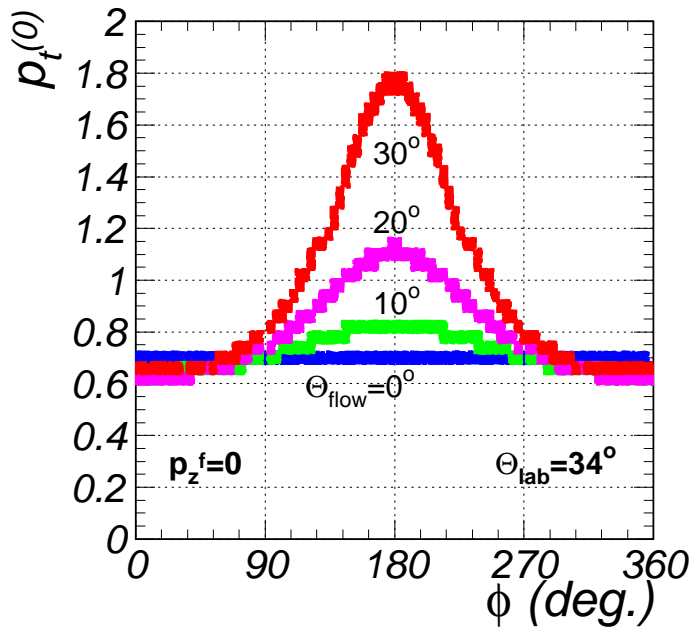, width=0.65\textwidth}}
\caption{The geometrical thresholds as function of the azimuthal angle in the
rotated reference frame for different flow angle values.} \label{fig-12}
\end{figure}

The FOPI device having an azimuthal symmetry \cite{gob,pet2,rit}, the border 
regions between different subdetectors are symmetric relative to the beam
axis. In other words, the transition between two consecutive subdetectors,
which causes small shadows (see Fig.~ref{fig-1}), 
takes place for a given rapidity at a given transverse momentum, and remains 
constant as a function of azimuth ($\phi$).
Once the reference frame is rotated, the symmetry is broken and the $\phi$ 
dependence of these regions have to be carefully treated. Fig.~\ref{fig-12}
shows as an example how the $\Theta_{lab}$=34$^\circ$ region (at midrapidity
in the rotated system, $p_z^f$=0) which corresponds to the borderline between 
CDC and forward Plastic Wall is seen in a $p_t^{(0)}-\phi$ representation in 
reference frames rotated relative to the beam axis by different 
$\Theta_{flow}$ values.

Due to this reason, for $p_t^{(0)}>$0.8, where the phase space is covered 
mainly by the CDC, the analyzed azimuthal range was [0$^\circ$,90$^\circ$] 
and [270$^\circ$,360$^\circ$]. The complete distributions have been
obtained by reflecting these ranges relative to 90$^\circ$ and 270$^\circ$,
respectively.

\section{Experimental Results}

In the present paper the azimuthal distributions are analyzed in a reference 
frame rotated in the reaction plane by $\Theta_{flow}$.
The midrapidity region is selected by the polar angular range in the
rotated reference frame $\Theta_{cm}^f$=80$^\circ$--100$^\circ$. 
For the present study, concentrated on the transition energy
such a selection is more adequate than one based on $p_z$ or rapidity.

\subsection{Flow angle} 

\begin{figure}[h]
\centering\mbox{\epsfig{file=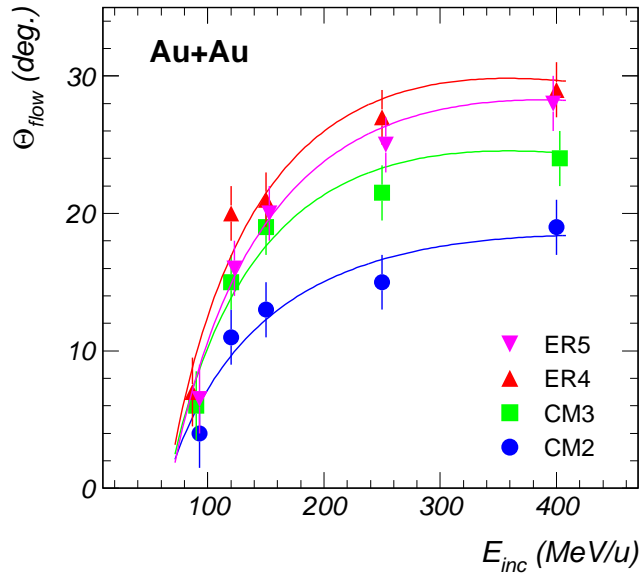, width=0.6\textwidth}}
\caption{The excitation function of the flow angle for the four centrality 
bins.}
\label{fig-9}
\end{figure}

The  incident energy dependence of $\Theta_{flow}$ determined 
using the recipe described in the previous chapter is shown in 
Fig.~\ref{fig-9} for different centralities.
The source of the slight differences relative to the ones obtained 
using transverse momentum analysis \cite{ram} has been discussed in the 
previous chapter.
An increase of $\Theta_{flow}$ as function of energy and centrality up to 
400$\cdot$A MeV and ER4, respectively, can be followed in Fig.~\ref{fig-9}. 
 The lines, drawn in order to guide the eyes, represent the result of 
polynomial fits to the experimental points.

As far as 90$\cdot$A MeV is in the range of $E_{tran}$ values (see later), 
the determination of $\Theta_{flow}$ using the present procedure is not very 
accurate. While this does not influence the results relative to $E_{tran}$ 
studies, it could bring a big uncertainty in the determination 
(by extrapolation) of the balance energy, $E_{bal}$. 
Although this subject will be properly treated in a separate
paper, it is worth to mention that the present analysis gives larger $E_{bal}$
values than those reported in ref. \cite{mag}, but in agreement with 
our previous results \cite{cro1}.

\subsection{General trends of the azimuthal distributions} 
For a multidimensional study of the incident energy at which a transition from
an in-plane enhancement to the out-of-plane preferential emission takes place, 
the excitation functions of azimuthal distributions for different fragment 
species, transverse momenta and centrality have been analyzed. 
We will present some of these trends in what follows.

\begin{figure}[hbt]
\begin{tabular}{lr} \begin{minipage}[t]{0.48\textwidth}
\centering\mbox{\epsfig{file=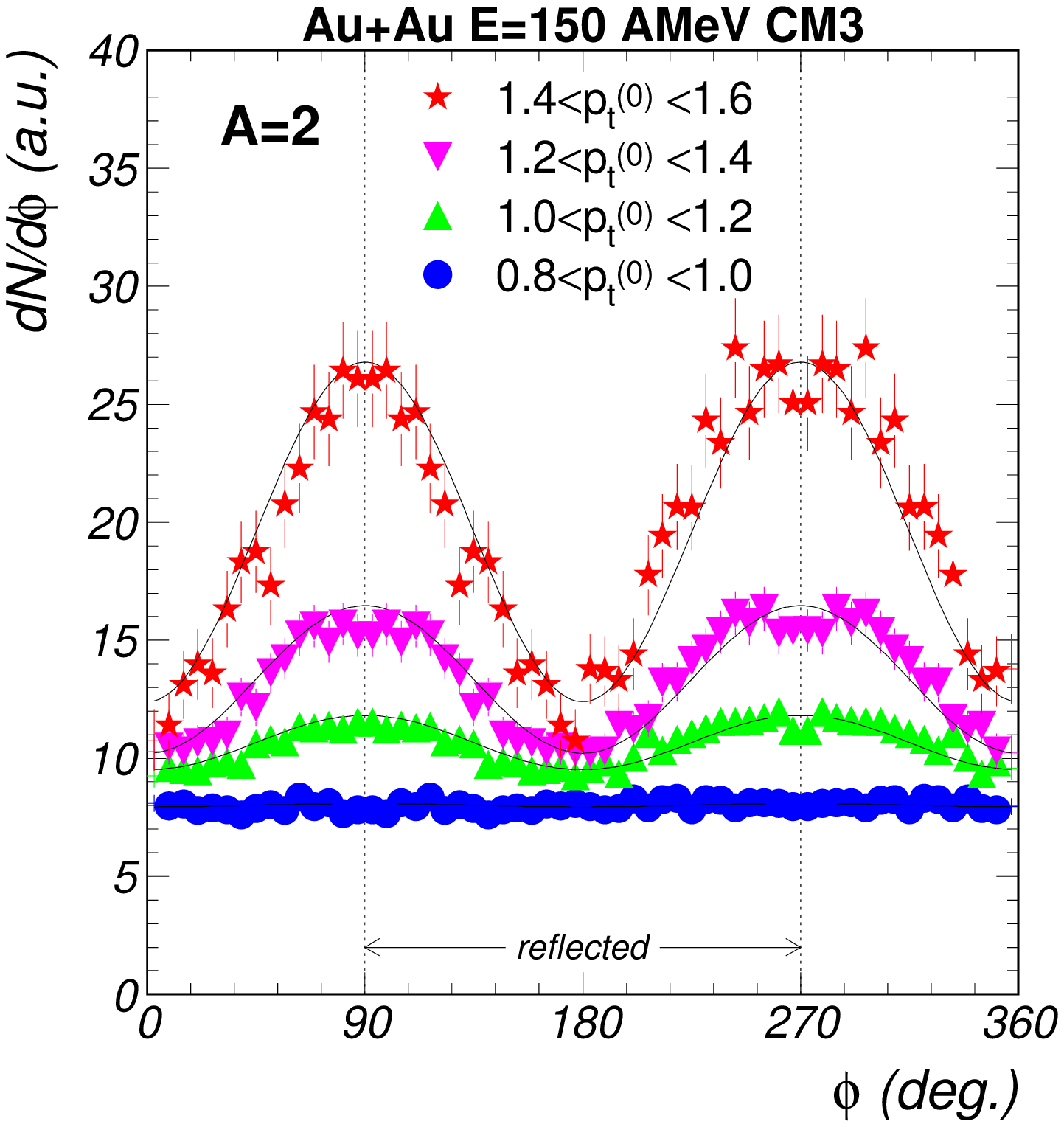,width=0.999\textwidth}}
\caption{The azimuthal distributions for different ranges of the scaled 
momentum $p_t^{(0)}$ for 150$\cdot$A MeV incident energy and A=2 
reaction products.} \label{fig-13}
\end{minipage} & \begin{minipage}[t]{0.48\textwidth}
\centering\mbox{\epsfig{file=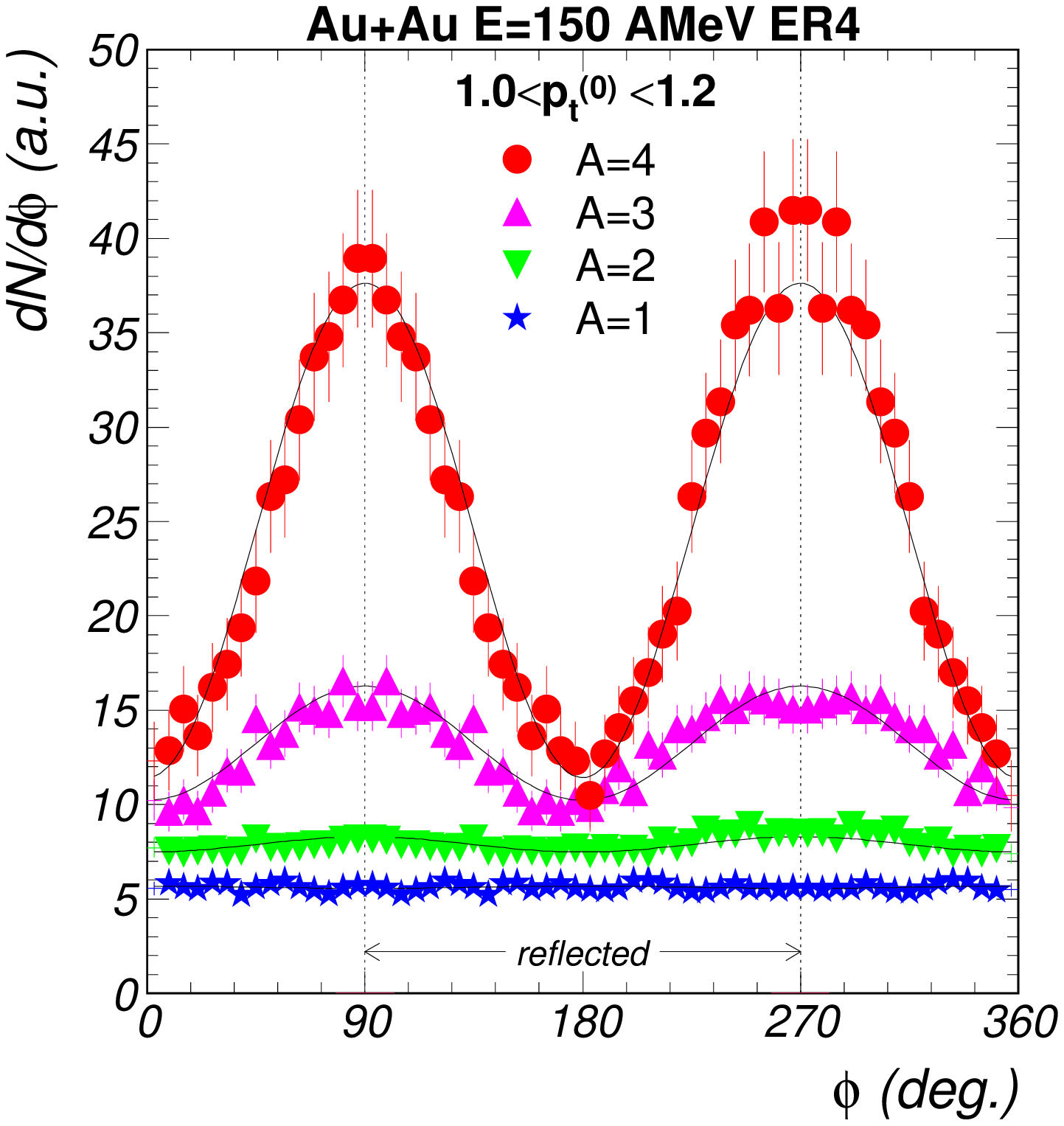,width=0.999\textwidth}}
\caption{The azimuthal distributions as function of the mass of the detected 
particle for 150$\cdot$A MeV and ER4 centrality.} \label{fig-14}
\end{minipage} \end{tabular}
\end{figure}

For a given incident energy (as an example 150$\cdot$A MeV was taken), the 
dependence of the azimuthal distributions as a function of the scaled 
transverse momentum $p_t^{(0)}$ is presented in Fig.~\ref{fig-13} for CM3 
centrality and A=2 reaction products. The enhancement of the anisotropy 
with increasing transverse momentum seen in the previous experiments 
\cite{lam,bri2,bas} is confirmed.
Taking the advantage of a complete phase space coverage of the present 
configuration of our experimental device, we chose to do the present studies 
for high transverse momentum $p_{t}^{(0)}$, a region where the theoretical 
predictions \cite{har1,bas2} suggested that the squeeze-out signal is
sensitive to the nuclear matter Equation of State (EoS).
For a given incident energy and range of $p_{t}^{(0)}$ the dependence of the 
azimuthal distribution as function of mass of the analyzed particle is shown
in Fig.~\ref{fig-14}.
The well known enhancement of the azimuthal anisotropy with increasing the 
mass of the analyzed reaction product is evidenced 
\cite{gut2,har1,bri2,wan,bas}. 

\begin{figure}[htb]
\begin{tabular}{lr} \begin{minipage}[t]{0.48\textwidth}
\centering\mbox{\epsfig{file=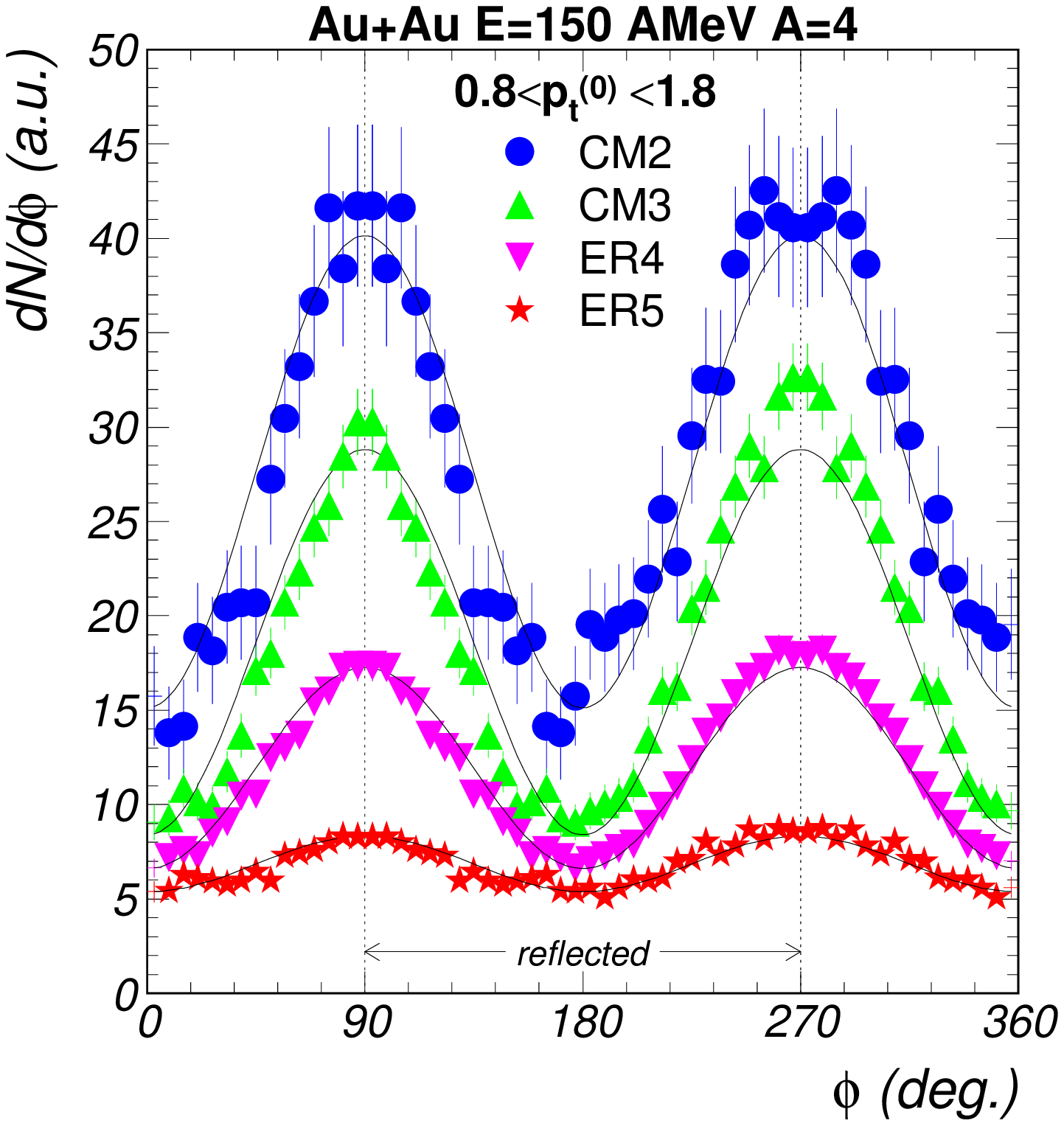,width=0.99\textwidth}}
\caption{Centrality dependence of the azimuthal distributions for 
150$\cdot$A MeV, A=4 fragments and 0.8$<p_{t}^{(0)}<$1.8 transverse momentum.} \label{fig-15}
\end{minipage} & \begin{minipage}[t]{0.48\textwidth}
\centering\mbox{\epsfig{file=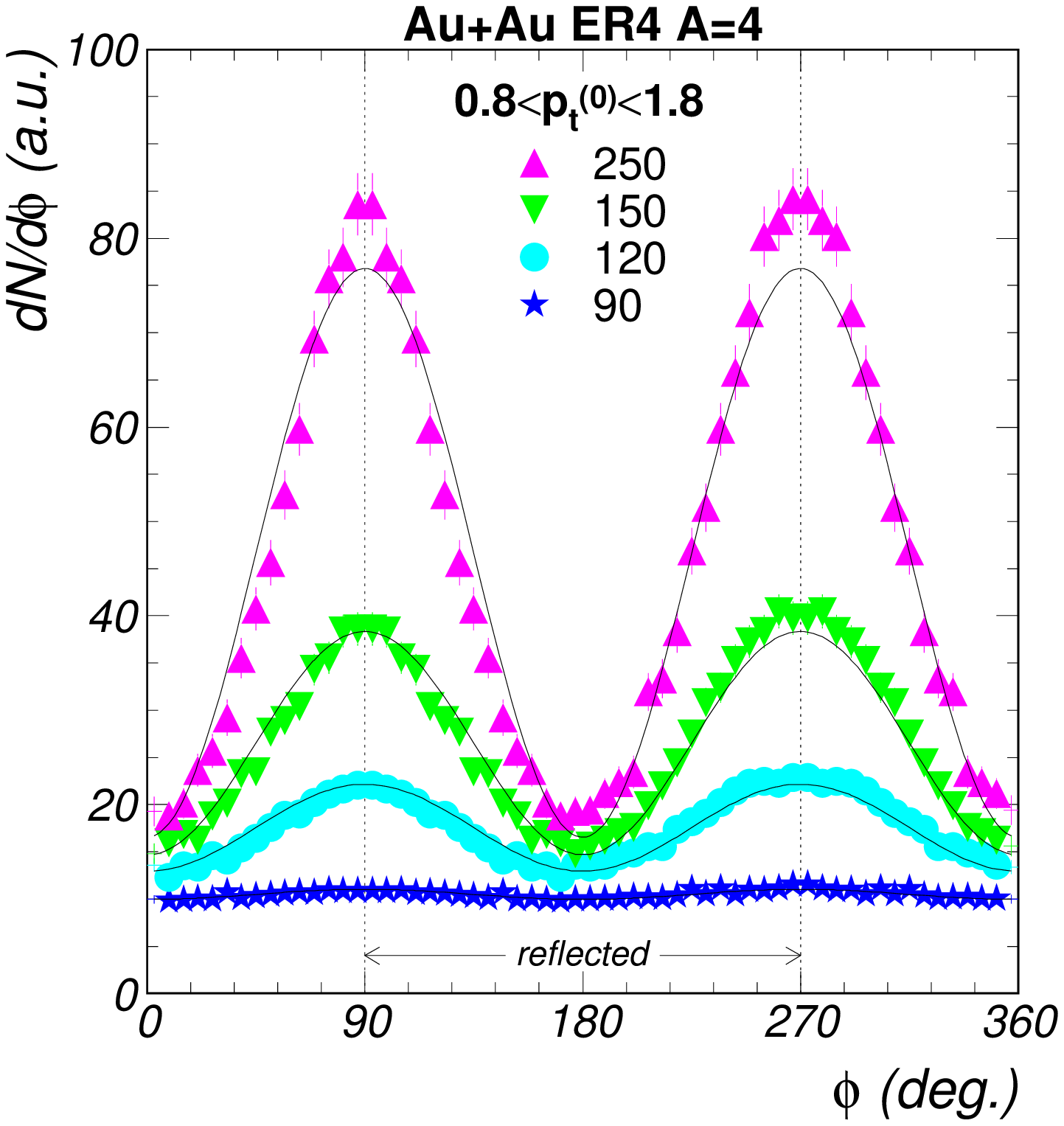,width=0.99\textwidth}}
\caption{Energy dependence of the azimuthal distributions for ER4 
centrality, A=4 fragments and 0.8$<p_{t}^{(0)}<$1.8 transverse momentum.} 
\label{fig-16} \end{minipage} \end{tabular}
\end{figure}

Fig.~\ref{fig-15} shows once more the importance of doing these types of 
studies in the appropriate coordinate frame \cite{gut1}. 
In the laboratory system the dependence of the squeeze-out signal as function
of the centrality is biased as far as one looks to the 
projection of these distributions in a plane which is not normal to the flow 
direction. As one can see from Fig.~\ref{fig-9} the flow angle increases 
with the centrality up to ER4 and this influences the information on the 
squeeze-out signal as a function of centrality if this information is 
extracted from the reference frame with the polar axis along 
the beam direction. This caution has to be taken when one studies the 
excitation function of the squeeze-out signal. 
A sample of the incident energy dependence of the azimuthal distributions for 
A=4 reaction products and for a given range in transverse momentum, 
0.8$<p_{t}^{(0)}<$1.8, is presented in Fig.~\ref{fig-16}.
All the azimuthal distributions presented are the primordial ones, namely 
not corrected for the reaction plane resolution.

As it was already mentioned, the azimuthal distributions have been analyzed 
in the rotated reference frame. They have been fitted with a second order 
Fourier expansion:
\begin{equation}
\left(\frac{dN}{d\phi}\right)^{exp} = 
a_0^{exp}\cdot(1 + a_1^{exp}\cos\phi + a_2^{exp}\cos2\phi)
\label{eq-2}
\end{equation}

 The smooth curves in Fig.~\ref{fig-13}-\ref{fig-16} have been obtained using 
this fit function. 
As there are some systematic dips around 90$\circ$ and 270$\circ$, caused by 
the detector acceptance in the rotated reference system and by the influence 
of the magnetic field on the reaction plane reconstruction, we excluded them
from the fits.
$a_{0}$ gives the average value of the distribution, 
$a_{1}$ corresponds to in-plane flow and $a_{2}$ describes the squeeze-out 
pattern. The squeeze-out ratio $R_{N}^{exp}$ can be written:
\begin{equation}
R_N^{exp}= \frac{1-a_2^{exp}}{1+a_2^{exp}}
\end{equation}

 The $a_{2}^{exp}$ values have been corrected for the fluctuations of the 
estimated reaction plane using the method of ref. \cite{oli} explained in the 
previous chapter:
\begin{equation}
a_2^{corr} = a_2^{exp}/\langle\cos 2\Delta\phi\rangle
\end{equation}
The corrected quantity $R_{N}$, which will be used in the following, 
is given by:
\begin{equation}
R_N= \frac{1-a_2^{corr}}{1+a_2^{corr}}
\end{equation}

\begin{figure}[hbt]
\centering\mbox{\epsfig{file=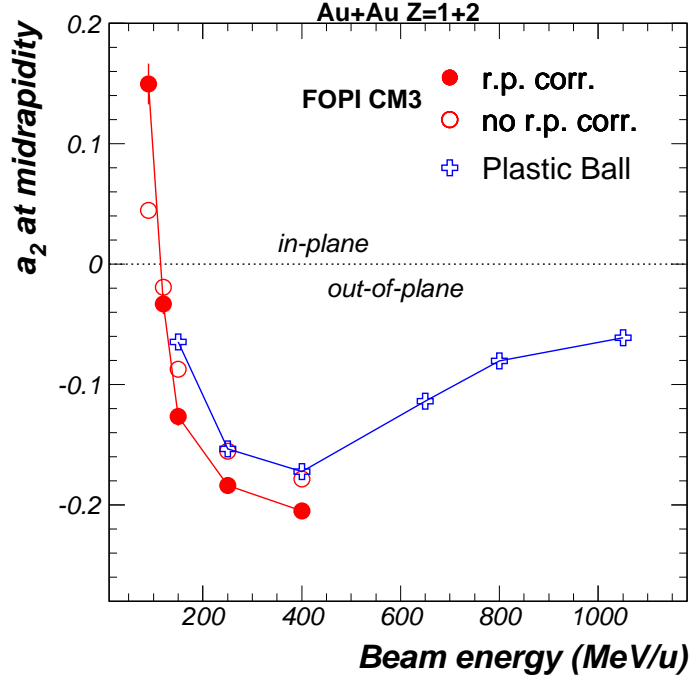,width=0.65\textwidth}}
\caption{The excitation function for the $a_{2}$ coefficient for the CM3
centrality (see text for details).}
\label{fig-24}
\end{figure}

Before proceeding to the detailed analysis of the transition energy, we 
present in Fig.~\ref{fig-24} the excitation function for the $a_{2}$ 
coefficient extracted using Z=1 and 2 particles and integrated over the 
transverse momentum. 
To compare with the Plastic Ball results \cite{gut2} which were not corrected 
for the reaction plane fluctuations, our $a_{2}$ values are shown 
both with (full dots) and without (open circles) this correction. Both 
sets of data are for roughly similar centrality bin (CM3 for the present 
analysis and MUL3 for the Plastic Ball data). 
Only for this case, we used a $p_z$ window to select the midrapidity,
as it was done in ref.~\cite{gut2}, namely $|(p_z^f/A)/(p_P^{cm}/A_P)|<$0.1, 
where $p_z^f/A$ is the particle longitudinal c.m. momentum per nucleon in 
the rotated reference frame and $p_P^{cm}/A_P$ is the projectile momentum 
per nucleon in the c.m. system.
A good agreement between the two data sets can be observed. This agreement is
meaningful as the resolutions of the reaction plane for the two experiments
are comparable, at least at 400$\cdot$A MeV \cite{gut2}.

The experimental $a_{2}^{exp}$ values in the rotated reference frame
at the incident energies, centralities, transverse momenta and for reaction 
species studied in this paper are given in the Appendix.

\vspace{5mm}
\subsection{Multidimensional analysis of the transition energy} 

 Most of the previous studies of the azimuthal distributions have been
concentrated on the behaviour of the observed anisotropies as a function of 
centrality, rapidity, transverse momentum and the mass of the reaction 
products. 
A detailed study of the incident energy at which a transition from in-plane to 
out-of-plane emission takes place could give deeper insight on the relative
contribution of the attractive and repulsive forces, the lifetime of the 
emitting source, its rotational energy and expansion dynamics.
 Earlier evidences of such a transition energy have been reported for Au+Au 
combination \cite{but1,aa1,bas,cro1} and Zn+Ni \cite{pop}. These studies have 
been done in the reference frame in which the polar axis was along the beam 
direction.
 At 100$\cdot$A MeV Au+Au collisions a transition from squeeze-out at high 
centrality to a rotational-like pattern towards peripheral collisions was
observed \cite{tsa1}.

\begin{figure}[htb]
\centering\mbox{\epsfig{file=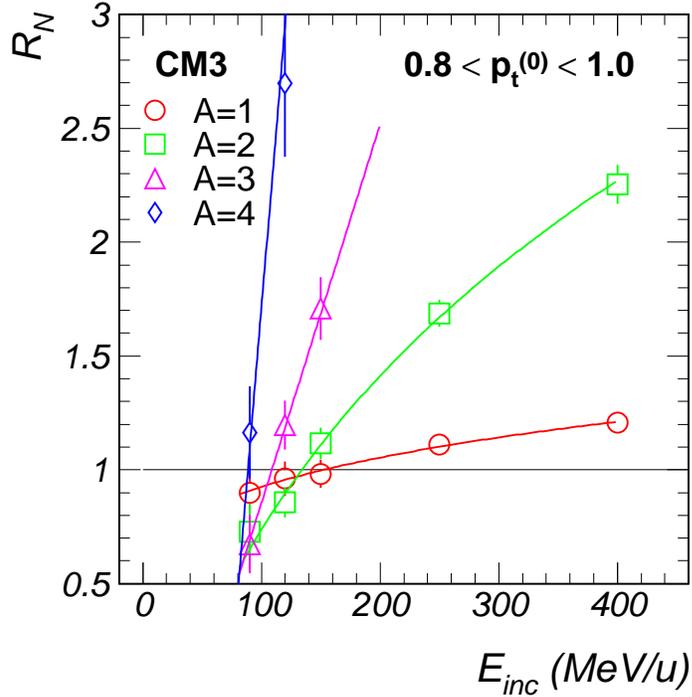,width=0.65\textwidth}}
\caption{The excitation function for the squeeze-out ratio for CM3
centrality and transverse momentum window 0.8$<p_{t}^{(0)}<$1.0.} 
\label{fig-17} \end{figure}

A multidimensional analysis of this transition energy in terms of its 
dependence as function of centrality, transverse momentum and mass of the 
reaction products for Au+Au is performed in this section.
 The procedure used to extract the transition energy ($E_{tran}$) can be 
followed in Fig.~\ref{fig-17}. For a given  
centrality range, transverse momentum and mass of the reaction 
product, the $R_{N}$ value is represented as a function of incident energy 
($E_{inc}$). 
Squeeze-out signal corresponds to $R_{N}>1$ while in-plane enhancement of the
azimuthal distribution is characterized by $R_{N}<1$. The continuous lines
correspond to the result of fits of the experimental points for different 
masses using a polynomial function.
The intersections of these lines with the one corresponding to 
$R_{N}=1$ value are defined as $E_{tran}$.  
Fig.~\ref{fig-17} shows a clear mass dependence of the transition energy, 
lower values corresponding to heavier particles for a given centrality and a 
given range in the scaled transverse momentum $p_{t}^{(0)}$.  

\begin{figure}[htb]
\centering\mbox{\epsfig{file=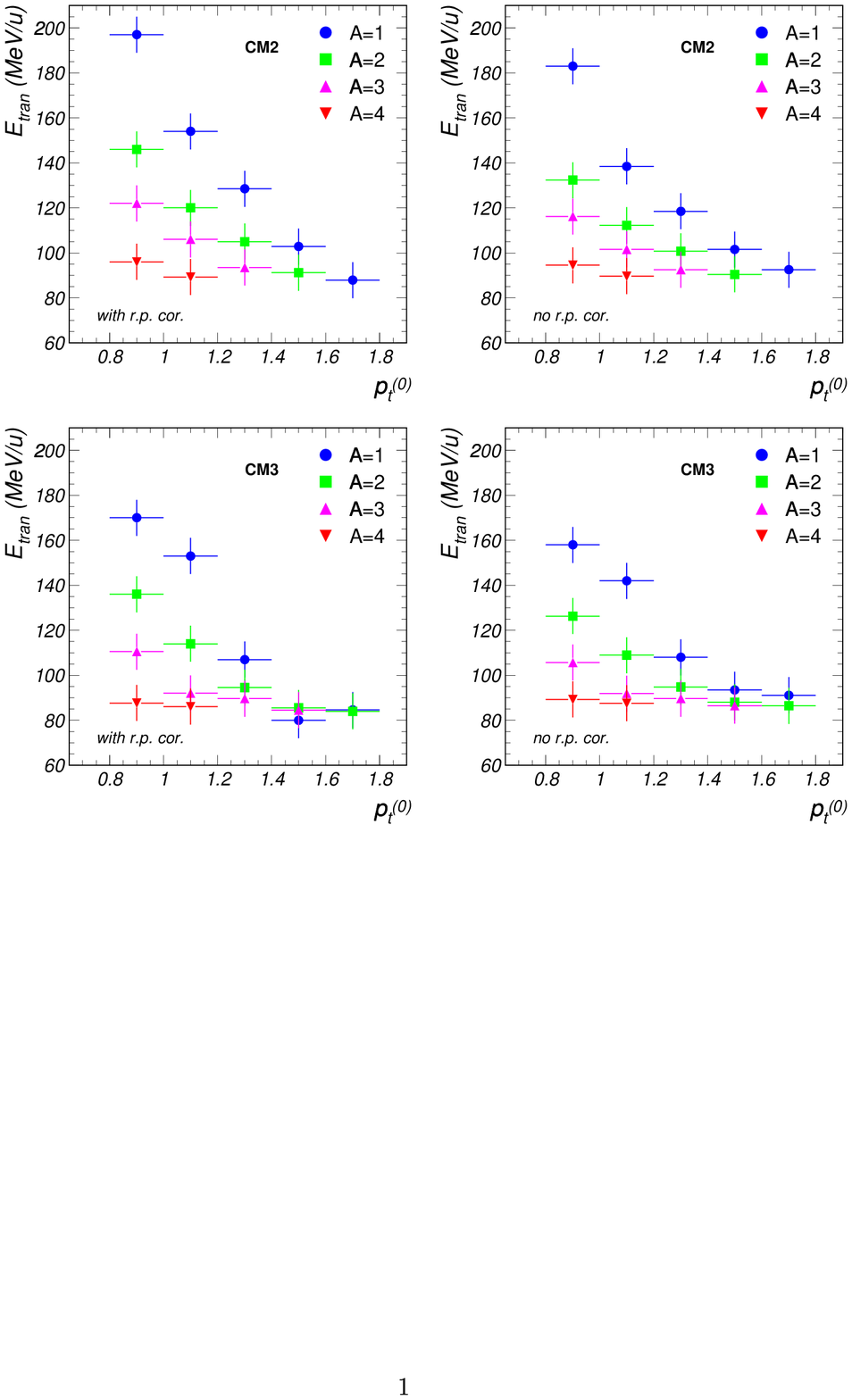,width=1.\textwidth}}
\caption{The transition energy as a function of the scaled momentum for 
different particle types for the CM2 and CM3 centrality bins, with (left 
column) or without (right column) reaction plane correction.} 
\label{fig-18}
\end{figure}

Following the above recipe for different regions of $p_{t}^{(0)}$, a two 
dimensional dependence of the $E_{tran}$ values as function of mass of the 
analyzed reaction products and transverse momentum can be obtained. 
The results are presented in Fig.~\ref{fig-18} for CM2 and CM3 centrality
bins, with and without doing the correction for the reaction plane resolution.
As expected, in case of using the correction, the values of $E_{tran}$ are 
systematically slightly higher, but the trends are identical for both cases.
A continuous decrease of the $E_{tran}$ as function of
$p_{t}^{(0)}$ is evidenced for all analyzed particles and the difference in
$E_{tran}$ values for different particles is decreasing towards larger values 
of $p_{t}^{(0)}$ for both centrality bins. 
One should mention here that for the cases when $R_{N}<1$ values have not 
been reached $E_{tran}$ was determined by extrapolating the fits of $R_{N}$ 
as function of incident energy ($E_{inc})$.

For a rotating emitting source one would expect a larger in-plane alignment 
for heavier fragments \cite{wil2}. 
This effect alone can not explain the mass dependence of 
$E_{tran}$. Therefore a dynamical effect has to be considered besides the 
pure geometrical one of shadowing. 
Different particles, originating from different regions of the fireball
\cite{pet1,kot} would feel the shadowing in a different way.
At large $p_t^{(0)}$ the contribution comes from larger expansion 
velocities, earlier expansion phase of the fireball \cite{pet1,pet2}
and consequently higher shadowing.
At lower values of $p_t^{(0)}$, the expansion zones are less localized, 
especially for the light fragments due to larger contribution of the thermal 
velocities relative to the collective ones, the shadowing being less effective
on the in-plane yields.
Contaminations coming from sequential evaporation processes smear-out such 
effects but cannot explain the observed trends.

\section{IQMD calculations} \label{s:model}

The IQMD (Isospin Quantum Molecular Dynamics) \cite{har3} is a QMD model 
\cite{aic} which takes into account the isospin degree of freedom for 
nucleon-nucleon cross section and Coulomb interaction. The main argument for 
using this version in the present context is based on the fact that IQMD 
has been successfully used for analyzing flow phenomena in heavy ion collisions
\cite{har1,har2,bas2,sof}. In this work two different parametrizations of 
the EoS are used, a hard EoS (compressibility $K$=~380 MeV) with 
momentum  dependence of the nucleon interaction (MDI) - HM and a soft EoS 
($K$=~200 MeV), with MDI - SM. 

The events produced by the model have been filtered by the 
experimental filter and analyzed in a similar way as the experimental data.
Table~\ref{tab-3} is the analog of Table~\ref{tab-1} but this time representing
the resolution of the reaction plane determination for the events generated by
this model (HM parametrization). 
The worse resolution at lower energies for calculated events is the
natural consequence of lower yield for intermediate mass fragments predicted by
the model. This was the reason why for the calculations we used the true 
reaction plane. 

\begin{table}[htb]
\begin{center}
\caption{The values of $\langle\cos\Delta\phi_R\rangle$ obtained 
analyzing the IQMD (HM) model events.}\label{tab-3}
\begin{tabular}{cccccc} \hline
Beam Energy  &90 &120 &150 & 250 & 400\\ \hline
CM2& ~~0.005 & ~~0.087 & ~~0.378 & ~~0.771 & ~~0.859 \\
CM3& ~~0.036 & ~~0.309 & ~~0.576 & ~~0.837 & ~~0.887 \\
ER4& ~~0.081 & ~~0.395 & ~~0.579 & ~~0.800 & ~~0.863 \\ 
ER5& ~~0.074 & ~~0.161 & ~~0.266 & ~~0.498 & ~~0.543\\ \hline
\end{tabular}
\end{center}
\end{table}

The model was used to estimate the accuracy of the reaction plane corrections
\cite{dan,oli} and it has been found that the correction used in the present 
study works fine in what concerns the sidewards flow, but it is overestimated 
for the squeeze-out signal, especially for the cases
$\langle\cos\Delta\phi_R\rangle<$0.4. We also found that applying the 
correction in the rotated frame slightly overestimates the real squeeze-out 
signal.

\begin{figure}[htb]
\centering\mbox{\epsfig{file=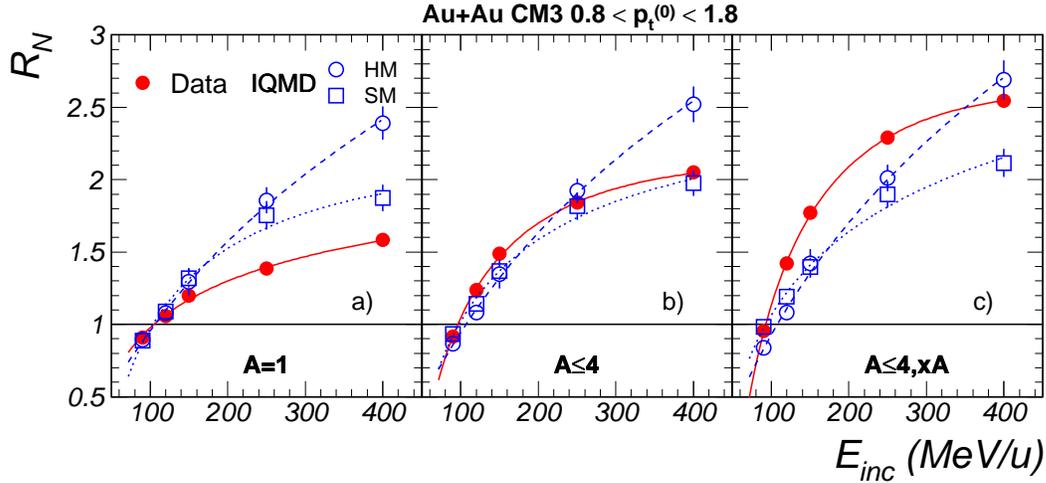, width=.99\textwidth}} %
\caption{The excitation function for the squeeze-out signal for CM3 centrality
bins, data and IQMD model for different particle selections.} \label{fig-22a}
\end{figure}

\begin{figure}[htb]
\centering\mbox{\epsfig{file=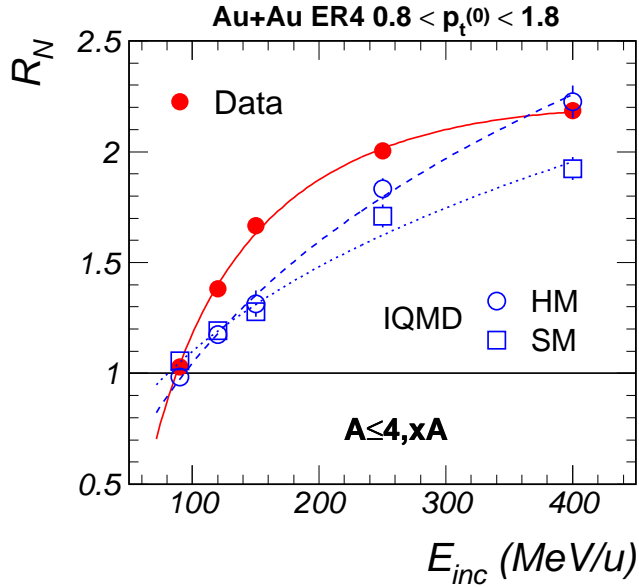, width=.6\textwidth}} %
\caption{The excitation function for the squeeze-out signal for ER4 centrality
bin, data and IQMD model for A$\le$4 particles weighted by their mass.} 
\label{fig-22} \end{figure}

As mentioned already, it is well known that the IQMD model underpredicts 
the composite particles yield \cite{tsa2,rei} and this drawback is especially 
important for our energy range. 
To establish the most meaningful way of comparison, in Fig.~\ref{fig-22a} 
we compare the excitation function of the squeeze-out ratio ($R_N$) for data 
and the model for three ways of particle selection: 
a) protons only, b) all particles up to alphas and c) a coalescence-type, 
summing and weighting the azimuthal distribution by each particle's atomic 
number $A$ (including up to alphas here as well).
It was shown within the IQMD model \cite{bas2} that the sensitivity of the 
squeeze-out signal to EoS is present only for high transverse momenta.
With our selection, $0.8<p_{t}^{(0)}<1.8$, at the higher energies we do see 
a dependence of the squeeze-out on the EoS.
Although the model predicts an increase of the squeeze-out signal with 
particle mass \cite{har1}, because the protons are dominating in IQMD
the model values for the three cases show only small variations.
This is clearly not the case with the data, where the inclusion of the 
composites, without or with weighting, is increasing the squeeze-out signal 
significantly.
For protons the model predicts larger squeeze-out than the experimental values
which compensates the lower yield of composite particles produced by a 
coalescence mechanism, so that integrating over A$\le$4 particles the agreement
with the experiment is good.
However, Fig.~\ref{fig-22a} shows that the nucleonic squeeze-out type of 
flow predicted by the model and the coalescence mechanism used to produce 
fragments do not explain the experimental observations.
This conclusion holds for the ER4 centrality as well. In Fig.~\ref{fig-22}
we present for this centrality the comparison for A$\le$4 particles weighted 
by their mass.
Apart of different magnitudes, the behavior is similar to the one presented
in Fig.~\ref{fig-22a}c.
All these facts show the importance of a comparison between the experiment 
and model either for each particle or for coalescence-type (as in 
Fig.~\ref{fig-22a}c and Fig.~\ref{fig-22}) and not simply summing-up all 
particles, when different effects could compensate each other and the 
agreement is artificially good.

Although at low energies the compressibility should indeed not play an
important role, there is a systematic difference between the two 
parametrizations, the HM version predicting a higher in-plane alignment.
Similar trends have been observed at 100$\cdot$A MeV for Au+Au by 
Aladin-Miniball Collaboration and comparisons with model predictions using 
BUU calculations show better agreement with a stiff EoS, momentum dependent 
nuclear interaction and reduced in-medium cross section 
($\sigma_{NN}=0.8\times\sigma_{free}$) \cite{tsa1}.
The in-plane enhancement in Ar+V system at energies below 100$\cdot$A MeV was 
studied using BUU calculations with $\sigma_{NN}=0.5\times\sigma_{free}$ and 
a stiff EoS \cite{wil2}.
More recently the transition region was investigated for Ca+Ca system using 
an isospin dependent BUU model and the dependences of $E_{tran}$ on 
EoS, MDI and $\sigma_{NN}$ were found to be significant \cite{zhe}.

\begin{figure}[htb] 
\centering\mbox{\epsfig{file=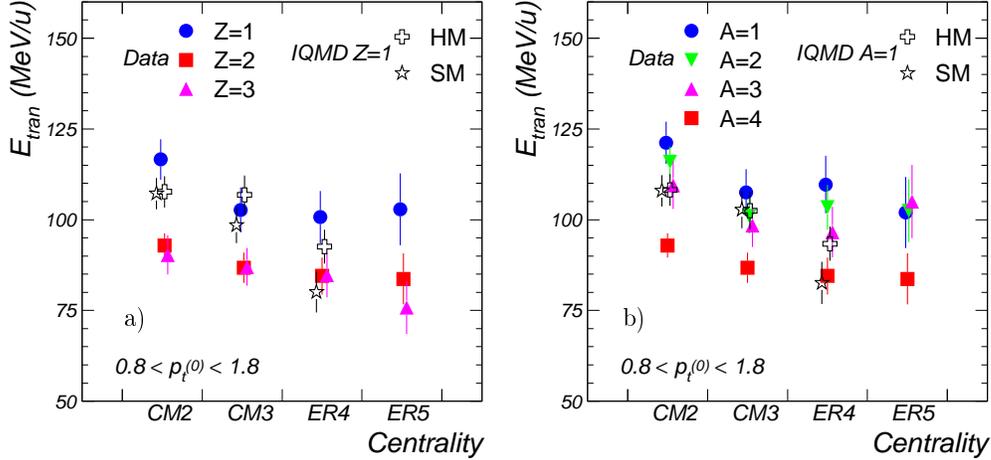, width=0.99\textwidth}}
\caption{Experimental transition energy values as function of centrality for:
a) Z=1,2,3 particles and b) A=1,2,3,4 particles. 
Full symbols denote the experimental values, open symbols are the IQMD model 
predictions: crosses - HM and stars - SM parametrization for the EoS.}
\label{fig-23}
\end{figure}

Although the number of generated events was about 2000 per 1 fm
interval in the impact parameter, the statistics for particles heavier than
A=1 was not sufficient to perform with reasonable accuracy a similar analysis
of $E_{tran}$ as the one done for the experimental data.
Consequently, we have chosen to do the comparison in a broad $p_{t}^{(0)}$
interval and as function of centrality.
The centrality dependence of the $E_{tran}$ values is presented in 
Fig.~\ref{fig-23}, for $0.8<p_{t}^{(0)}<1.8$ window and particles selected
according to: a) charge (Z=1-3) and b) mass (A=1-4). 
For Z=1 and A=1, respectively, the experimental data are compared with 
the IQMD model with the two versions for compressibility. 
The shown centrality bins (ER5, ER4, CM3 and CM2) correspond to the impact
parameter intervals 0-2, 2-4, 4-6 and 6-8 fm, respectively, as it was 
explained in the second chapter of this paper. 
Systematically, the SM version predicts lower transition energies relative 
to HM version.
The model predicts an evolution of the $E_{tran}$ as function of 
centrality for Z=1 and A=1 particles similar to the experiment. 
With decreasing impact parameter the amount of energy which can be 
transferred in a rotational-like motion becomes smaller. 
At the same time the volume of the spectator matter causing the shadowing 
is reduced, while the number of baryons involved in the fireball region, 
$A_{part}$, increases, contributing to larger expansion 
\cite{dan3,lis,pak,aa2}. 
All these effects play a role in the observed experimental azimuthal 
distributions. However, the centrality dependence of the $E_{tran}$ suggests 
that the reduction of the rotational motion of the fireball is the main 
source of the observed effect.
One could observe in Fig.~\ref{fig-23} a less pronounced dependence on 
centrality for Z=1 particles relative to the heavier ones. 
A possible explanation would be that at higher centralities, the
three sources of emitting particles, the projectile spectator, the target 
spectator and the participant zone, are less separated, the mixed 
contribution being enhanced for lighter reaction products.
The contribution from sequential processes brings an additional smearing of 
anisotropies, increasing artificially the $E_{tran}$ value for lighter 
particles.

\section{Conclusions}

Using a new generation of data collected with the complete FOPI experimental
device we presented a systematic study of the excitation function of the
azimuthal distributions around midrapidity. For Au+Au system for which the 
most complete excitation function was measured, starting with 90$\cdot$A MeV,
a clear transition from in-plane (rotational-like) to out-of-plane 
(squeeze-out) was evidenced to take place going from lower to higher energies.
The incident energy at which this transition takes place, called $E_{tran}$, 
was studied as function of mass of reaction products, their scaled transverse 
momentum ($p_{t}^{(0)}$) and centrality. Special attention was payed to the 
sidewards flow angle ($\Theta_{flow}$) determination used to define the 
rotated reference frame in which all these studies have been done. 

In general one could say that a systematic decrease of the $E_{tran}$ values 
with increasing the mass of the analyzed fragment, the transverse momentum 
and centrality of the collision was observed.
$E_{tran}$ values being the incident energies where the effect of the rotation
of a compound system created by the mean field, expansion of a hot and 
compressed participant zone built-up by many consecutive nucleon-nucleon 
interactions and the shadowing effect of the colder spectator regions 
compensate each other, it seems to be an experimental observable which could 
distinguish between different theoretical models and different approximations 
used by them.
Studying an expanding-rotating hot and compressed fireball in the 
presence of the shadowing objects can give additional information on the 
expansion dynamics. The observed decrease of the $E_{tran}$ as a function of 
the mass of the emitted fragments can be the result of a complex dynamics,
indicating that heavier particles for which the thermal contribution
is less important are better localized in the expansion process using 
$p_{t}^{(0)}$ selection than the lighter ones. 
When integrating over a large range in momenta, our $E_{tran}$ values are
in a good agreement with previous studies
which found the transition to take place around 100$\cdot$A MeV.

Detailed comparisons with the IQMD model predictions evidenced the necessity
to avoid any kind of fragment mixture in order to draw significant conclusions.
Squeeze-out ratio for protons and coalescence-type revealed that nucleonic 
out-of-plane flow and the coalescence mechanism used by the model to create
complex fragments can not explain the experimental observations. 
The low statistics of IQMD generated events did not allow a similar analysis 
of $E_{tran}$ as for the experiment. 
Integrating on $p_{t}^{(0)}$, the systematic decrease of
$E_{tran}$ values with increasing the collision centrality is reproduced 
by the model. Although systematically the hard EoS gives larger $E_{tran}$
values, closer to the experimental values, the small difference relative 
to the soft EoS is not sufficient to draw definite conclusions on the EoS. 
    
\begin{ack}
This work has been supported in part by the German BMBF under contracts 
RUM-005-95, POL-119-95, UNG-021-96 and RUS-676-98 and by the Deutsche 
Forschungsgemeinschaft (DFG) under projects 436 RUM-113/10/0, 
436 RUS-113/143/2 and 446 KOR-113/76/0,
Support has also been received from the Polish State Committee of Scientific 
Research, KBN, from the Hungarian OTKA under grant T029379 and from the 
Korea Research Foundation under contract No. 1997-001-D00117.
\end{ack}

\vspace{3mm}
{\normalsize \bf Appendix: Experimental $a_2$ values}

\vspace{3mm}
In this Appendix we present the values of the $a_{2}^{exp}$ 
coefficient for different particle types and ranges in normalized transverse
momenta. Remind that these values are the ones obtained from the fit of the
experimental data with expression \ref{eq-2}, namely without including the 
correction for the reaction plane resolution.
In parentheses we quote the absolute errors on the last digit of the
respective number.

\vspace{5mm}
\begin{table}[hbt]
\begin{center}
\caption{The $a_{2}^{exp}$ values in the rotated reference frame for 
particles with A=1, 2 and 3, for CM2 centrality bin.} \label{tab-a}

\vspace{5mm}
\begin{tabular}{ccccccc} \hline
$p_{t}^{(0)}$ & A & 90 & 120     &  150     &  250     & 400 \\ \hline
0.8-1.0 & 1 & 0.034(9)& 0.044(9) & 0.020(9) & -0.030(6)& -0.110(6)\\ 
        & 2 & 0.07(1) & 0.06(1)  & -0.02(1) & -0.230(7)& -0.400(8)\\ 
        & 3 & 0.08(1) & 0.01(1)  & -0.12(1) & -0.48(1) & -0.71(1)\\ \hline
1.0-1.2 & 1 & 0.047(8)& 0.030(9) & -0.004(9)& -0.110(6)& -0.210(7)\\ 
        & 2 & 0.06(1) & 0.00(1)  & -0.12(1) & -0.400(9)& -0.59(1)\\ 
        & 3 & 0.05(2) & -0.09(2) & -0.31(2) & -0.71(2) & -0.87(2)\\ \hline
1.2-1.4 & 1 & 0.044(9)& 0.013(9) & -0.060(9)& -0.180(7)& -0.310(8)\\ 
        & 2 & 0.04(1) & -0.07(2) & -0.24(2) & -0.55(1) & -0.71(2)\\ 
        & 3 & 0.03(2) & -0.27(4) & -0.55(5) & -0.88(2) & -0.81(6)\\ \hline
1.4-1.6 & 1 & 0.03(1) & -0.050(9)& -0.12(1) & -0.310(8)& -0.44(1)\\ 
        & 2 & 0.01(2)& -0.24(3) & -0.40(3) & -0.72(2) & -0.89(3)\\ \hline
\end{tabular}
\end{center}
\end{table}

\begin{table}[hbt] 
\begin{center}
\caption{Same as Table~\ref{tab-a}, but for CM3 centrality bin.} \label{tab-b}

\vspace{5mm}
\begin{tabular}{ccccccc} \hline
$p_{t}^{(0)}$ & A & 90 & 120     &  150     &  250     & 400 \\ \hline
0.8-1.0 & 1 & 0.018(9)& 0.022(9) & 0.007(9) & -0.030(6)& -0.060(6)\\ 
        & 2 & 0.04(1) & 0.04(1)  & -0.04(1) & -0.200(6)& -0.310(7)\\ 
        & 3 & 0.06(1) & -0.04(1) & -0.18(1) & -0.45(1) & -0.62(1)\\ \hline
1.0-1.2 & 1 & 0.021(8)& 0.025(9) & -0.001(9)& -0.070(6)& -0.140(6)\\ 
        & 2 & 0.05(1) & -0.01(1) & -0.12(1) & -0.340(7)& -0.470(9)\\ 
        & 3 & -0.00(2)& -0.15(2) & -0.35(2) & -0.63(2) & -0.77(2)\\ \hline
1.2-1.4 & 1 & 0.023(8)& -0.010(9)& -0.060(9)& -0.140(7)& -0.230(8)\\ 
        & 2 & 0.01(1) & -0.11(2) & -0.26(2) & -0.48(1) & -0.60(1)\\ 
        & 3 & -0.01(2)& -0.29(3) & -0.51(3) & -0.78(3) & -0.89(3)\\ \hline
1.4-1.6 & 1 & 0.014(9)& -0.040(9)& -0.14(1) & -0.230(8)& -0.35(1)\\ 
        & 2 & -0.03(2)& -0.20(2) & -0.39(2) & -0.62(2) & -0.71(2)\\
        & 3 & -0.10(5)& -0.49(6) & -0.60(6) & -0.8(2) & - \\ \hline
\end{tabular}
\end{center}
\end{table}

\begin{table}[hbt] 
\begin{center}
\caption{Same as Table~\ref{tab-a}, but for ER4 centrality bin.} \label{tab-c}

\vspace{5mm}
\begin{tabular}{ccccccc} \hline  
$p_{t}^{(0)}$ & A & 90& 120      &  150     &  250     & 400 \\ \hline
0.8-1.0 & 1 & 0.021(9)& 0.034(9) & 0.012(9) & -0.026(6)& -0.068(6)\\ 
        & 2 & 0.04(1) & 0.04(1)  & -0.05(1) & -0.19(1) & -0.24(1)\\ 
        & 3 & 0.04(1) & -0.04(1) & -0.12(1) & -0.32(1) & -0.48(1)\\ \hline
1.0-1.2 & 1 & 0.019(8)& 0.037(9) & 0.008(9) & -0.068(6)& -0.12(1)\\ 
        & 2 & 0.04(1) & -0.02(1) & -0.09(1) & -0.24(1) & -0.37(1)\\ 
        & 3 & -0.02(2)& -0.10(2) & -0.30(2) & -0.49(1) & -0.61(2)\\ \hline
1.2-1.4 & 1 & 0.014(8)& 0.018(9) & -0.052(9)& -0.13(1) & -0.21(1)\\ 
        & 2 & 0.02(2) & -0.06(2) & -0.17(2) & -0.42(1) & -0.49(1)\\ 
        & 3 & -0.05(3)& -0.22(3) & -0.35(3) & -0.66(2) & -0.75(2)\\ \hline
1.4-1.6 & 1 & 0.029(9)& -0.074(9)& -0.09(1) & -0.20(1) & -0.26(1)\\ 
        & 2 & -0.04(2)& -0.20(2) & -0.25(2) & -0.52(2) & -0.58(2)\\ \hline
\end{tabular}
\end{center}
\end{table}

\begin{table}[hbt] 
\begin{center}
\caption{Same as Table~\ref{tab-a}, but for ER5 centrality bin.} \label{tab-d}

\vspace{5mm}
\begin{tabular}{ccccccc} \hline  
$p_{t}^{(0)}$ & A & 90& 120      &  150     &  250     & 400 \\ \hline
0.8-1.0 & 1 & 0.03(1) & 0.02(2)  & 0.00(2)  & -0.01(2) & -0.03(1)\\ 
        & 2 & 0.03(1) & 0.03(1)  & -0.01(2) & -0.09(2) & -0.15(1)\\ 
        & 3 & 0.02(1) & 0.00(2)  & -0.05(2) & -0.11(1) & -0.23(1)\\ \hline
1.0-1.2 & 1 & 0.00(1) & 0.01(1)  & -0.01(2) & -0.04(2) & -0.05(2)\\ 
        & 2 & 0.01(1) & 0.00(1)  & -0.04(2) & -0.12(2) & -0.21(2)\\ 
        & 3 & 0.00(2) & -0.04(2) & -0.13(2) & -0.25(2) & -0.34(2)\\ \hline
1.2-1.4 & 1 & 0.01(1) & -0.03(2) & -0.05(2) & -0.08(2) & -0.11(2)\\ 
        & 2 & 0.01(2) & -0.03(2) & -0.11(2) & -0.19(2) & -0.28(2)\\ 
        & 3 & 0.01(2) & -0.08(2) & -0.14(2) & -0.36(2) & -0.39(4)\\ \hline
\end{tabular}
\end{center}
\end{table}

\vspace{5mm}
\begin{table}[hbt]
\begin{center}
\caption{The $a_{2}^{exp}$ values in the rotated reference frame for 
particles with Z=1, 2 and 3, integrated over the momentum interval 
\mbox{0.8$<p_{t}^{(0)}<$1.8}.} \label{tab-e}

\vspace{5mm}
\begin{tabular}{ccccccc} \hline  
Centrality & Z & 90 & 120      &  150      &  250      & 400 \\ \hline
CM2 & 1 & 0.040(4) & -0.022(4) & -0.084(4) & -0.240(3) & -0.330(3)\\ 
    & 2 & 0.02(1)  & -0.23(2)  & -0.43(2)  & -0.82(2)  & -0.93(2)\\ 
    & 3 & 0.01(1)  & -0.37(4)  & -0.64(5)  & -0.84(4)  & - \\ \hline
CM3 & 1 & 0.022(4) & -0.034(3) & -0.120(3) & -0.241(2) & -0.298(3)\\ 
    & 2 & -0.04(1) & -0.33(1)  & -0.56(1)  & -0.80(1)  & -0.89(1)\\ 
    & 3 & -0.06(2) & -0.43(2)  & -0.72(3)  & -0.97(3)  & - \\ \hline
ER4 & 1 & 0.017(6) & -0.029(5) & -0.10(4) & -0.190(3)  & -0.240(3)\\ 
    & 2 & -0.05(1) & -0.29(1)  & -0.49(1)  & -0.68(1)  & -0.78(1)\\ 
    & 3 & -0.09(3) & -0.40(2)  & -0.61(2)  & -0.84(2)  & -0.88(4)\\ \hline
ER5 & 1 & 0.008(6) & -0.010(9) & -0.05(1)  & -0.10(1)  & -0.14(1)\\ 
    & 2 & -0.02(1) & -0.15(2)  & -0.29(2)  & -0.43(2)  & -0.54(2)\\ 
    & 3 & -0.06(2) & -0.18(5)  & -0.36(4)  & -0.48(4)  & -0.63(5)\\ \hline
\end{tabular}
\end{center}
\end{table}

\vspace{5mm}
\begin{table}[hbt] 
\begin{center}
\caption{The $a_{2}^{exp}$ values in the rotated reference frame for 
particles with A=1,2,3 and 4, integrated over the momentum interval 
\mbox{0.8$<p_{t}^{(0)}<$1.8}.} \label{tab-f}

\vspace{5mm}
\begin{tabular}{ccccccc} \hline 
Centrality & A & 90 & 120      &  150      &  250      & 400 \\ \hline
CM2 & 1 & 0.034(6) & -0.001(4) & -0.050(4) & -0.170(6) & -0.260(6)\\ 
    & 2 & 0.052(6) & -0.010(6) & -0.110(6) & -0.361(4) & -0.50(1)\\ 
    & 3 & 0.06(1)  & -0.05(1)  & -0.19(1)  & -0.55(1)  & -0.74(2) \\
    & 4 & 0.02(1)  & -0.23(2)  & -0.43(2)  & -0.82(2)  & -0.93(3) \\ \hline
CM3 & 1 & 0.017(6) & -0.013(4) & -0.070(8) & -0.160(4) & -0.221(4)\\ 
    & 2 & 0.025(6) & -0.05(1)  & -0.15(1)  & -0.34(1)  & -0.43(1)\\ 
    & 3 & 0.03(1)  & -0.10(1)  & -0.27(2)  & -0.52(1)  & -0.66(1) \\ 
    & 4 & -0.04(1) & -0.33(2)  & -0.56(2)  & -0.80(2)  & -0.89(2) \\ \hline
ER4 & 1 & 0.018(8) & -0.007(9) & -0.061(8) & -0.110(6) & -0.160(6)\\ 
    & 2 & 0.023(8) & -0.03(1)  & -0.11(1)  & -0.27(1)  & -0.33(1)\\ 
    & 3 & 0.00(1)  & -0.06(1)  & -0.20(2)  & -0.39(1)  & -0.52(1)\\ 
    & 3 & -0.05(1) & -0.29(2)  & -0.49(2)  & -0.68(2)  & -0.78(2) \\ \hline
ER5 & 1 & 0.006(6) & -0.005(9) & -0.027(9) & -0.044(8) & -0.079(8)\\ 
    & 2 & 0.011(8) & -0.03(1)  & -0.05(1)  & -0.13(1)  & -0.20(1)\\ 
    & 3 & 0.01(1)  & -0.04(2)  & -0.09(2)  & -0.17(1)  & -0.29(2)\\
    & 4 & -0.02(1) & -0.15(1)  & -0.29(1)  & -0.43(1)  & -0.54(2) \\ \hline
\end{tabular}
\end{center}
\end{table}

\end{document}